\documentclass[a4paper,11pt]{article}
\pdfoutput=1 

\usepackage{jheppub} 
\usepackage{hyperref}
\usepackage[T1]{fontenc} 
\usepackage{amsmath,amssymb}
\usepackage{slashed}
\usepackage{tikz}
\usepackage{upgreek}
\usetikzlibrary{arrows,positioning,cd,calc,math,decorations.pathreplacing,decorations.markings,decorations.pathmorphing
}

\tikzset{wave/.style={decorate, decoration=snake}}

\newcommand{\tr}{\text{tr\,}}

\newcommand{\Tau}{\mathcal{T}}
\newcommand{\res}{\text{Res\,}}
\newcommand{\diag}{\text{diag}}

\newcommand{\bs}{\boldsymbol}
\newcommand{\mc}{\mathcal}
\newcommand{\eq}[1]{\begin{equation}\begin{gathered}#1\end{gathered}\end{equation}}

\title{Circular quiver gauge theories, isomonodromic deformations and $W_N$ fermions on the torus}

\author[a,b,c]{Giulio Bonelli,}
\author[a,b,c]{Fabrizio Del Monte,}
\author[d,e,f]{Pavlo Gavrylenko}
\author[a,b,c]{and\\ Alessandro Tanzini}

\affiliation[a]{International School of Advanced Studies (SISSA),
via Bonomea 265, 34136 Trieste, Italy}
\affiliation[b]{Institute for Geometry and Physics (IGAP),
via Beirut 2/1, 34151 Trieste, Italy}
\affiliation[c]{INFN, Sezione di Trieste}
\affiliation[d]{Center for Advanced Studies, Skolkovo Institute of Science and Technology, Nobel Street 1, 121205 Moscow, Russia}
\affiliation[e]{NRU HSE, International Laboratory of Representation Theory and Mathematical Physics, Usacheva 6, 119048 Moscow, Russia}
\affiliation[f]{Bogolyubov Institute for Theoretical Physics, Metrologichna 14-b,  03143 Kyiv, Ukraine}

\emailAdd{bonelli@sissa.it}
\emailAdd{fdelmont@sissa.it}
\emailAdd{pasha.145@gmail.com}
\emailAdd{tanzini@sissa.it}

\abstract{
We study the relation between 
class $\mathcal{S}$ theories on punctured tori and isomonodromic deformations of flat $SL(N)$ connections on the two dimensional torus with punctures.
Turning on the self dual $\Omega$-background 
corresponds to a deautonomization of the Seiberg-Witten integrable system
which implies a specific time dependence in its Hamiltonians.
We show that the corresponding $\tau$-function is proportional to 
the dual gauge theory partition function, the proportionality factor being 
a non trivial function of the solution of the deautonomized Seiberg-Witten integrable system.
This is obtained by mapping the isomonodromic deformation problem to 
$W_N$ free fermion correlators on the torus.
}

\begin{document} 

\maketitle
\flushbottom

\section{Introduction}\label{sec:Intro}

A major problem in modern Quantum Field Theory is that of 
understanding its non perturbative formulation.
While this issue is somehow accessible in low space-time dimensions ($d=0,1,2$), in higher dimensions this turned out to be achievable only for
particular classes of models, namely supersymmetric ones.
In these cases, due to a sophisticated analysis of the quantum measure and of the 
Feynman path-integral, it is possible to perform exact computations of BPS saturated sectors of 
the theory \cite{Pestun:2016zxk}, that are reduced essentially to matrix models.
A crucial aspect of these results is that special functions and transcendental functions show up as basic building blocks.
This is indeed an expected feature from several general view points, first of all
from the analysis of the asymptotic nature of the power series in the coupling constants arising in perturbative QFT \cite{Dyson:1952tj}.

A particular set of results in this wider framework, were started by the analysis
of \cite{Gamayun:2013auu}, where a link between Painlev\'e transcendents and 
multi-instanton counting in ${\mathcal N}=2$ $d=4$ $SU(2)$ SUSY gauge theories in self-dual $\Omega$-background \cite{Nekrasov:2003af} was noticed.
Further analysis has shown the natural identification to be between partition functions and solutions of Painlev\'e equations in $\tau$-form.

This was not the first time in which Painlev\'e transcendents arose in gauge theory. Indeed Painlev\'e functions show up already in $d=0$ gauge theory, namely matrix models. The most famous example appears in the analysis 
of the Hermitian matrix model with cubic potential of Kontsevich and 
Painlev\'e I equation \cite{Gross:1989aw}. As an important fact, the full matrix model partition function has been identified with the KP $\tau$-function
in \cite{Kharchev:1991cu}. 

\begin{figure}\label{Fig:CircularQuiver}
\begin{center}
\begin{tikzpicture}

\foreach \ang\lab\anch in {90/N/north,  0/N/east, 270/N/south, 180/N/west}{
  \draw[fill=orange,thick] ($(0,0)+(\ang:3)$) circle (.5);
  \node[anchor=\anch] at ($(0,0)+(\ang:3.3)$) {$\lab$};
}

\foreach \ang\lab in {90/N,180/N}{
  \draw[->,thick,shorten <=17pt, shorten >=17pt] ($(0,0)+(\ang:3)$) arc (\ang:\ang-90:3);
}

 \node at ($(0,0)+(45:3.5)$) {$m_1$};
  \node at ($(0,0)+(180-45:3.5)$) {$m_n$};

\draw[->,thick,shorten <=17pt] ($(0,0)+(0:3)$) arc (360:325:3);
\draw[->,thick,shorten >=17pt] ($(0,0)+(305:3)$) arc (305:270:3);
\draw[->,thick,shorten <=17pt] ($(0,0)+(270:3)$) arc (270:235:3);
\draw[->,thick,shorten >=17pt] ($(0,0)+(215:3)$) arc (215:180:3);
\node at ($(0,0)+(0-25:3.5)$) {$m_2$};
\node at ($(0,0)+(270+25:3.5)$) {$m_{i-1}$};
\node at ($(0,0)+(270-25:3.5)$) {$m_i$};
\node at ($(0,0)+(180+25:3.5)$) {$m_{n-2}$};

\foreach \ang in {310,315,320,220,225,230}{
  \draw[fill=black] ($(0,0)+(\ang:3)$) circle (.02);
}

\end{tikzpicture}
\end{center}
\caption{Circular quiver gauge theory corresponding to the torus with $n$ punctures: for every puncture $z_i$ we have a hypermultiplet of mass $m_i$ sitting in the bifundamental representation of two  different $SU(N)$ gauge groups. The case $n=1$ is special, as the hypermultiplet is in the bifundamental representation for the same $SU(N)$ gauge group, so that it is an adjoint hypermultiplet and the theory is the $\mathcal{N}=2^*$ theory.}
\end{figure}
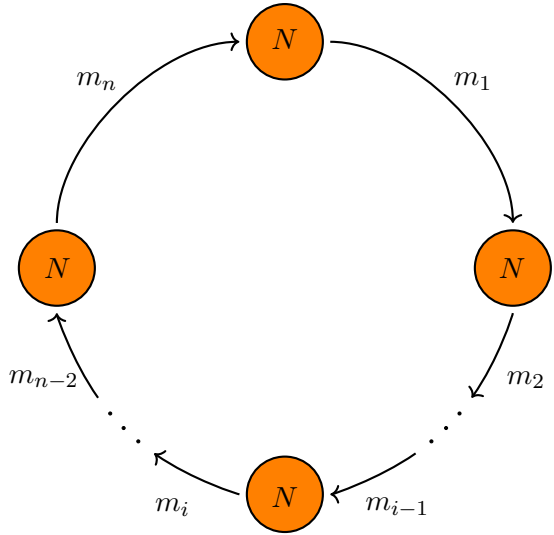

In this paper we will analyse how the identification between gauge theory partition function and the $\tau$-function of a suitable isomomonodromy deformation problem (of which Painlev\'e equations constitute the simplest instance) arises for a $A_{N-1}$ class $\mathcal{S}$ theories on the torus, a typical example of which is a circular quiver ${\mathcal N}=2$ $d=4$ $SU(N)$ SUSY gauge theory, depicted in Figure \ref{Fig:CircularQuiver}, in a 
self-dual $\Omega$-background and which are the integrable systems involved, generalizing the result of \cite{Bonelli:2019boe}, where the simplest of such theories, namely the $SU(2)$ $\mathcal{N}=2^*$ gauge theory, was shown to be related to the elliptic form of the Painlev\'e VI equation \cite{manin1996sixth}. 

In order to understand the correspondence between isomonodromy deformations and four-dimensional $\mathcal{N}=2$ supersymmetric gauge theories a central object is the Hitchin system \cite{Hitchin:1987mz}, in terms of which it is possible to formulate Seiberg-Witten theory, describing the Coulomb branch of the theory \cite{Donagi:1995cf}. The appearance of such an object is best understood within the context of class $\cal S$ theories \cite{Gaiotto:2009we,Gaiotto:2009hg,Bonelli:2009zp,Teschner:2010je,Bonelli:2011na}: one obtains theories in this class by compactifying the $A_{N-1}$ six dimensional $(2,0)$ superconformal field theory on a Riemann surface $\Sigma_{g,n}$ of genus $g$ with $n$ punctures, with punctures carrying also additional information given by singular boundary conditions for the fields. The basic reason for the appearance of Hitchin systems is that the four dimensional theory preserves $\mathcal{N}=2$ supersymmetry iff the internal fields $(A,\phi)$ on $\Sigma_{g,n}$ satisfy Hitchin equations:
\begin{equation}\label{eq:Hitchin}
\begin{cases}
F+[\phi,\bar{\phi}]=0, \\
\bar{\partial}\phi=0,
\end{cases}
\end{equation}
with singular behavior at the punctures specified by the boundary conditions. On the one hand, the moduli space of these equations is a hyperk\"ahler manifold given by the total space of a torus fibration, whose base space can be identified with the Coulomb branch of the four-dimensional gauge theory; on the other hand, this space is known to be an algebraic integrable system. In the $I$ complex structure the Hitchin system reduces to a Higgs bundle  whose spectral curve
\begin{equation}
\Sigma_{SW}:\det(\phi-\lambda)=0,
\end{equation}
can be identified with the Seiberg-Witten curve. The "Higgs field" $\phi$ of the Higgs bundle defined by \eqref{eq:Hitchin} is the Lax matrix of the integrable system. The question of how this picture gets modified when one tries to follow the physics from the deep IR of the Coulomb branch was asked since the early days of Seiberg-Witten theory, and the answer to this question was found to be that one has to split the times of the integrable system into "slow" and "fast" times, effectively deautonomizing the system in a consistent way: this corresponds, in the language of integrable system, to the so-called Whitham deformations \cite{Gorsky:1995zq,Edelstein:1999xk}. However, the procedure to study Whitham deformations is very involved, and with this method it is only possible to reconstruct the physics order by order in the deformation.

Actually it turns out to be more convenient to start from the UV physics, which is described by instanton counting in terms of Nekrasov partition functions \cite{Nekrasov:2003af,Nekrasov:2003rj}. First, an expression for the tau function of isomonodromic deformations of Higgs bundles corresponding to theories in class $\cal S$ associated to Riemann surfaces of genus zero has been found as a Fourier transform of Virasoro \cite{Gamayun:2012ma,Iorgov:2014vla} or $W_N$ conformal blocks \cite{Gavrylenko:2015wla,Gavrylenko:2018ckn} on the sphere. By using the AGT correspondence \cite{Alday:2009aq} one can show that this object is essentially identified with the Nekrasov-Okounkov dual partition function (modulo some known proportionality factor) for linear quiver gauge theories in class $\cal S$ \cite{Gamayun:2013auu,Bonelli:2016qwg}:
\begin{equation}\label{eq:TauZD1punct}
Z^D\propto\Tau,
\end{equation}
where the Nekrasov-Okounkov dual partition function is a discrete Fourier 
transform of the full Nekrasov partition function with respect to the Cartan parameters.
In the cases where we have only one isomonodromic flow and two-dimensional monodromy manifold, the deformation equations are Painlev\'e equations, and the degeneration of one Painlev\'e equation into another is precisely mapped to the decoupling of hypermultiplets in the corresponding gauge theory, or in some cases to taking the limit to an Argyres-Douglas fixed point \cite{Bonelli:2011aa,Nagoya:2015cja,Bonelli:2016qwg,Lisovyy:2018mnj}. These equations have the natural interpretation of exact, nonperturbative renormalization group equations for the asymptotically free gauge theories, since the deformation times are given by the dynamically generated scale, or as conformal manifold equations, since in the conformal case the times are given by the exactly marginal deformations of the theory. This picture has been generalized in \cite{Bonelli:2019boe} to the case of Gaiotto curves of genus one by considering the specific case of the torus with one puncture. This corresponds to the ${\mathcal N}=2^*$ gauge theory with 
$SU(2)$ gauge group.
It was there shown that beyond genus zero the above picture is slightly modified, since now
\begin{equation}\label{eq:TauSL2}
\Tau=\frac{\eta^2(\tau)}{\theta_1(Q(\tau))^2}Z^D_{\mathcal{N}=2^*},
\end{equation}
where $\eta(\tau)$ is Dedekind's eta function and $Q(\tau)$ solves the particular case of Painlev\'e VI equation in elliptic form \cite{manin1996sixth}
\begin{equation}
(2\pi i)^2\frac{d^2Q}{d\tau^2}=m^2\wp'(2Q).
\end{equation}
Thus in this case the proportionality factor is not just a simple function, but a highly transcendental one.

These results show that the integrable structure underlying the UV theory (in the self-dual omega background) is not the Higgs bundle itself, but rather its isomonodromic deformation, which corresponds to the oper limit of the Hitchin system in the complex structure $J$ \cite{Bonelli:2016qwg}. This realizes in an exact way the original idea of using Whitham deformation to describe the physics outside of the deep IR regime. Further, given the appearance not just of Nekrasov partition functions, but rather of their dual version, it seems more natural to reformulate the CFT solution by using free fermion conformal blocks \cite{Gavrylenko:2016moe}, that naturally yield a Fourier series structure from the sum over fermionic charges in the Fock space. In fact, it was shown in \cite{Bonelli:2019boe} that as soon as one goes beyond genus zero, this reformulation is not just more natural, but actually necessary, so that it really seems the correct framework for this problem.

Another feature of working with free fermions is that there is a natural connection of free fermions with the theory of topological strings, where they appear in various contexts \cite{Aganagic:2003qj,Marino:2011eh,Dijkgraaf:2008fh,Bonelli:2016idi,Bonelli:2017ptp,Coman:2018uwk}. The topological strings in turn engineer theories of class $\cal S$ when formulated on certain toric Calabi-Yau manifolds \cite{Katz:1996fh,Hollowood:2003cv}. In fact, it turns out that these isomonodromic deformations underlie topological strings only in the geometric engineering limit, where we have a theory of class $\cal S$. The full topological string partition function itself, as computed with the (unrefined) topological vertex \cite{Aganagic:2003db}, are instead related to tau function of $q$-Painlev\'e equations \cite{Bonelli:2017gdk,Bershtein:2017swf,Bershtein:2018srt,Mironov:2019pij}, and $q$-Virasoro conformal blocks \cite{Bershtein:2016aef,2017arXiv170601940J,Matsuhira:2018qtx,Mironov:2017sqp}. In fact, the connection with isomonodromy problems goes beyond the perturbative setting of the topological vertex, making contact with the nonperturbative proposal of \cite{Grassi:2014zfa} for the Topological String partition function (see also \cite{Grassi:2019coc} for recent developments).

Our main result is the generalization of \eqref{eq:TauSL2} to the case of an $SL(N,\mathbb{C})$ isomonodromic problem on the torus with an arbitrary number of punctures, i.e. to class $\cal S$ circular quiver theories with $SU(N)$ gauge groups obtained by compactifying the six-dimensional $A_{N-1}$ $(2,0)$ superconformal field theory on a genus one surface with $n$ regular punctures at positions $z_1,\dots,z_n$. We show that the expression \eqref{eq:TauSL2} is generalized to this case in the following way:
\begin{equation}\label{eq:ZDTauIntro}
\Tau=Z^D\prod_i\frac{\eta(\tau)}{\theta_1(Q_i(\{z_k\},\tau)-\sigma\tau-\rho)},
\end{equation}
where $Q_i$ are again the dynamical variables of the isomonodromic system. These solve a system of coupled nonlinear differential equations corresponding to an elliptic version of the Schlesinger system, in which the times are the punctures' positions $z_1,\dots,z_n$ and the elliptic modulus $\tau$ \cite{Levin:2013kca}, and $Z^D$ is a Fourier transform of free fermionic conformal blocks, of the form
\begin{equation}\label{eq:ZDIntro}
Z^D=\tr_{\mathcal{H}}\left(q^{L_0}(-)^Fe^{2\pi i\bs\eta\cdot\bs{J_0}} V_1\dots V_n\right).
\end{equation}
$\bs{J_0}$ are charges under the Cartan of a twisted $\widehat{\mathfrak{gl}(N)}_1$ algebra and $\bs\eta$ their fugacities. $\sigma,\rho$ are the $U(1)$ charge and fugacity of this $\widehat{\mathfrak{gl}(N)}_1$. When the vertex operators $V_1\dots V_n$ are semi-degenerate fields of $W_N$, through the AGT correspondence $Z^D$ is identified with the dual partition function of a circular quiver gauge theory, while for more general values of their W-charges the derivation, while formally holding at the level of CFT, does not have a known gauge theory counterpart, and thus an explicit combinatorial expression in terms of Nekrasov functions \cite{Nekrasov:2003af,Nekrasov:2003rj,Bruzzo:2002xf}. Note that while the representation \eqref{eq:ZDIntro} corresponds to the dual partition function of a circular quiver gauge theory, by applying fusion transformations on the vertex operators it is possible to obtain the other class $\cal S$ theories corresponding to the same number of punctures on the torus. The corresponding tau functions will differ by connection constants determined by the fusion kernels, as it happens in the case of the sphere \cite{Iorgov:2013uoa,Its:2014lga,Its:2016jkt,Lisovyy:2018mnj}. The construction contains additional $U(1)$ parameters over which the tau function does not depend. We show however that the zeroes of the dual partition function in these additional variables are exactly the solutions $Q_i$ of the nonautonomous system. The condition
$Z^D=0$
is therefore shown to be
the nonautonomous generalization of the algebro-geometric solution of the Calogero-Moser model \cite{1999JMP....40.6339G}. Moreover, the fact that the tau function does not depend on $\sigma,\rho$ can be made explicit by decomposing the trace in \eqref{eq:ZDIntro} into different $\mathfrak{sl}_N$ sectors, labeled by $j=1,\dots,N$. We can then rewrite the relation \eqref{eq:ZDTauIntro} as\footnote{See section \ref{sec:EllGaudin} for details.}
\begin{equation}
Z^D_j=\frac{\Theta_j(\textbf{Q})}{\eta(\tau)^{N-1}}\Tau,
\end{equation}
where now $Z^D_j$, $j=0,\dots,N-1$ are $N$ different dual partition functions for the $SU(N)$ quiver theory, with different shifts in the Fourier series over the Coulomb branch parameters, and $\Theta_j$ are Riemann theta functions given by equation \eqref{eq:RiemannTheta}.

The paper is structured as follows: in Section \ref{sec:LinSys} we define the rank $N$ isomonodromic problem on the torus with $n$ regular singularities; in Section \ref{sec:FreeFerm} we introduce $N$-component complex free fermions and their related vertex operators, that we then use in Section \ref{sec:FreeFermSol} to provide an expression for the kernel and tau function of the isomonodromic problem. In section \ref{sec:VerlindeSol} we provide an alternative proof of the statements of section \ref{sec:FreeFermSol} by using the technique of Verlinde loop operators: while less general than that of the preceding section, this proof has the upside of being less formal, and to every quantity is provided an explicit expression. In section \ref{sec:KricheverCFT} we discuss the CFT solution to the linear system defined by Krichever's approach to isomonodromic deformations \cite{Krichever:2001cx}, which we also briefly discuss. In section \ref{sec:EllGaudin} we use our results to find an explicit formula for the solutions of the elliptic Schlesinger system as zeros of the dual partition function, generalizing the algebro-geometric solution of the Elliptic Calogero-Moser integrable system \cite{1999JMP....40.6339G} to the nonautonomous case with arbitrary number of singular points. 
In Appendix \ref{sec:Theta} we provide our notations about elliptic and theta functions, while in Appendix \ref{sec:WNApp} we briefly recall some generalities about $W_N$ algebras and their (semi-) degenerate fields.

\section{General Fuchsian system on the torus}\label{sec:LinSys}
We are going to study monodromy preserving deformations of linear systems on the torus of the form
\begin{equation}\label{eq:LinSys}
\partial_z Y(z|\tau)=L(z|\tau)Y(z|\tau), \\
\end{equation}
where $L,Y$ are $N\times N$ matrices and $L$, the Lax matrix, has $n$ simple poles located at $\{z_1,\dots,z_n\}$, also called Fuchsian singularities. Differently from what happens on the sphere, $L(z)dz$ is not a single-valued matrix differential, but rather has the following twist properties along the torus A- and B-cycles \cite{Korotkin:1995yi,Korotkin:1999xx,Takasaki:2001fr,Levin:2013kca}:
\begin{align}\label{eq:LaxTwist}
L(z+1)=T_AL(z)T_A^{-1}, && L(z+\tau)=T_BL(z)T_B^{-1}.
\end{align}
As can be seen from \eqref{eq:LinSys}, these twists will act on the solution $Y$ of the linear system on the left, in addition to the usual right-action by monodromies. Note that, while the monodromies are left invariant by the isomonodromic flows $z_1,\dots,z_n,\tau$, this is not true for the twists. In fact, as was already discussed in \cite{Bonelli:2019boe}, the twists are essentially parametrized by the dynamical variables of the isomonodromic system, of which $z_1,\dots,z_n,\tau$ are the times. 
The analytic continuation of $Y$ along the generators $\gamma_1,\dots,\gamma_n,\gamma_A,\gamma_B$ of $\pi_1(\Sigma_{1,n})$ is then
\begin{equation}\label{eq:Monodromies}
\begin{cases}
Y(\gamma_k\cdot z|\tau)=Y(z|\tau)M_k, \\
Y(z+1|\tau)=T_A(\{z_i\},\tau)Y(z|\tau)M_A, \\
Y(z+\tau|\tau)=T_B(\{z_i\},\tau)Y(z|\tau)M_B.
\end{cases}
\end{equation}
Together with the singular behavior of $Y$ around $z_1,\dots,z_n$, which are its branch points, these conditions fix completely $Y(z|\tau)$.

As discussed in \cite{Levin:2013kca}, for the group $SL(N,\mathbb{C})$ there are $N$ inequivalent Lax matrices of this kind characterized by the commutation relation of the twists:
\begin{equation}
T_AT_B^{-1}T_A^{-1}T_B=e^{2\pi ic_1/N},
\end{equation}
where $c_1=0,\dots,N-1$ is the first Chern class of the bundle having the centre of $SL(N,\mathbb{C})$ as structure group. It is possible to relate Lax matrices characterizing inequivalent bundles by means of singular gauge transformations, called Hecke modifications of the bundle \cite{Levin:2001nm}. Another possible approach, as in \cite{Krichever:2001cx}, is to consider instead a single-valued Lax matrix with additional simple poles at the so-called 'Tyurin points'. We will discuss the CFT solution to the problem defined by this latter Lax matrix, and its relation to our approach, in Section \ref{sec:KricheverCFT}.

Because of $\eqref{eq:Monodromies}$, it is possible to define the following kernel:
\begin{equation}\label{eq:KernelDef}
K(z',z)\equiv Y^{-1}(z')\Xi(z',z)Y(z),
\end{equation}
where $\Xi$ is defined so that it has one simple pole at $z=z'$, and transforms as
\begin{align}
\Xi(z'+1,z)=T_A\Xi(z',z), && \Xi(z',z+1)=\Xi(z',z)T_A^{-1},
\end{align}
\begin{align}
\Xi(z'+\tau,z)=T_BM_B^{U(1)}\Xi(z',z), && \Xi(z',z+\tau)=\Xi(z',z)\left(M_B^{U(1)}\right)^{-1}T_B^{-1},
\end{align}
in such a way that its transformation cancels the twists of $Y$. We also included the possibility for $\Xi$ to introduce further $U(1)$ factors, which will be useful to compare with the free fermion description. Because of this, along a closed cycle $\gamma$, $K$ transforms as follows
\begin{align}\label{eq:KernelMonodromy}
K(\gamma\cdot z',z)={\hat M}^{-1}_\gamma K(z',z), && K(z',\gamma\cdot z)=K(z',z){\hat M}_\gamma,
\end{align}
where
\begin{equation}
\hat M_\gamma=M_\gamma M_\gamma^{U(1)}
\end{equation}
is the $GL_N$ representative of $\gamma$ in the monodromy group, while $M_\gamma$ is its $SL_N$ representative (the monodromy of the solution $Y$). 

Keeping in mind the aforementioned fact that we can straightforwardly change from one bundle to another by means of a (singular) gauge transformation, from now on we consider the case $c_1=0$ of a topologically trivial bundle, for which the Lax matrix has the form
\begin{equation}\label{eq:LaxTorus}
L(z|\tau)=\textbf{p}+ \sum_{k=1}^nL^{(k)}, 
\end{equation}
where
\begin{equation}
\textbf{p}=\diag(p_1,\dots,p_N)
\end{equation}
and
\begin{equation}
L_{ij}^{(k)}=\delta_{ij}\frac{\theta_1'(z-z_k)}{\theta_1(z-z_k)}S^{(k)}_{ii}+
(1-\delta_{ij})\frac{\theta_1'(0)\theta(z-z_k-Q_i+Q_j)}{\theta_1(z-z_k)\theta_1(-Q_i+Q_j)}S_{ji}^{(k)},
\end{equation}
where the parameters $S_{ii}^{(k)}$ are subject to the constraint
\begin{equation}
\sum_k S_{ii}^{(k)}=0,
\end{equation}
so that we have the correct quasi-periodicity properties \eqref{eq:LaxTwist}. The monodromy preserving deformations of \eqref{eq:LinSys} involve moving the singular points $z_1,\dots,z_k$ (one of which can be fixed using the automorphisms of the torus), and the modular parameter $\tau$. These flows are generated by the Hamiltonians, given by the trace  of the Lax matrix squared
\begin{equation}\label{eq:LaxHamilt1}
\frac{1}{2}\tr L^2(z)=H_\tau+\sum_{k=1}^n H_k E_1(z-a_k)+C_2^k E_2(z-a_k),
\end{equation} 
where $E_1,E_2$ are the Eisenstein functions (see Appendix \ref{sec:Theta} for their definition), $C_2^k$ is the Casimir at the orbit of $z_k$, while $H_k,H_\tau$ generate the flows with times $z_k$ and $2\pi i\tau$ respectively, and can be computed by performing contour integrals:
\begin{align}\label{LaxHamilt2}
H_k=\oint_{\gamma_k}\frac{dz}{2\pi i}\frac{1}{2}\tr L^2(z), && H_\tau=\oint_Adz\frac{1}{2}\tr L^2(z).
\end{align}
These Hamiltonians can all be obtained as usual from the logarithmic derivative of a single tau function \cite{Korotkin:1995yi,Korotkin:1999xx,Takasaki:2001fr,Levin:2013kca}:
\begin{align}\label{eq:TauHamilt}
\partial_{z_k}\log\Tau=H_k, && 2\pi i \partial_\tau\log\Tau=H_\tau.
\end{align}

\section{$N$-component free fermions}\label{sec:FreeFerm}

In \cite{Bonelli:2019boe} it was shown that to describe $SL(2,\mathbb{C})$ isomonodromic deformations on the torus it is not sufficient to consider representations of Virasoro algebra, but we have to extend our space to include also a Fock space $\mathfrak{F}_\sigma$ with vacuum charge $\sigma$. This generalizes to the $SL(N,\mathbb{C})$ case by considering representation of $W_N$ algebra, rather than Virasoro, as in \cite{Gavrylenko:2015wla,Gavrylenko:2018ckn}. In turn, this makes contact, rather than with the usual AGT correspondence \cite{Alday:2009aq}, with a four-dimensional limit of topological strings, that are more naturally connected to free fermions \cite{Aganagic:2003qj,Dijkgraaf:2007sw,Dijkgraaf:2008fh,Coman:2018uwk}. Due to the extra Fock space, the system that is needed in the end is that of $N$-component complex free fermions, which we define in this section without introducing degenerate fields of $W_N$. The more "traditional" approach to isomonodromy involving degenerate fields and Verlinde loop operators is described in Section \ref{sec:VerlindeSol}.

The approach that we will adopt is very close to that of \cite{Gavrylenko:2016moe}: we define $N$-component free complex fermions, collecting them in two vectors $\psi,\bar{\psi}$, by their Fourier expansion in cylindrical coordinates:
\begin{align}
\psi(z)=\sum_{r\in\mathbb{Z}+1/2}\psi_r e^{2\pi i(r+\bs a+\frac12)z}, && \bar{\psi}(z)=\sum_{r\in\mathbb{Z}+1/2}\bar{\psi}_re^{2\pi i(r-\bs a-\frac12)},
\end{align}
or in components
\begin{align}
\psi_\alpha(z)=\sum_{r\in\mathbb{Z}+1/2}\psi_{\alpha,r} e^{2\pi i(r+a_\alpha+\frac12)z}, && \bar{\psi}_\alpha(z)=\sum_{r\in\mathbb{Z}+1/2}\bar{\psi}_{\alpha,r}e^{2\pi i(r-a_\alpha-\frac12)}
\end{align}
Here $\bs a$ is in the Cartan of $\mathfrak{sl}_N$, and the Fourier modes of the components $\psi_\alpha(z)$, $\psi_\beta(z)$ satisfy the usual canonical anticommutation relations
\begin{align}
\{\psi_{\alpha,r},\psi_{\beta,s} \}=\{\bar{\psi}_{\alpha,r},\bar{\psi}_{\beta,s}\}=0, && \{\bar{\psi}_{\alpha,r},\psi_{\beta,s} \}=\delta_{\alpha,\beta}\delta_{r,-s}, \\ r,s\in\mathbb{Z}+1/2, && \alpha,\beta=1,\dots,N.
\end{align}
The fermionic bilinear operators
\begin{equation}
J_{\alpha\beta}(z)\equiv :\bar{\psi}_\alpha(z)\psi_\beta(z):
\end{equation}
generate a twisted $\widehat{\mathfrak{gl}(N)}_1$ algebra, whose Cartan subalgebra can be used to define a $W_N\otimes\mathfrak{F}$ subalgebra. Its generators are given as elementary symmetric polynomials of the Cartan currents:
\begin{equation}
W_n(z)\equiv\sum_{\alpha_1<\dots<\alpha_n}:J_{\alpha_1}\dots J_{\alpha_n}:
\end{equation}
where $n=1,\dots,N$, and
\begin{align}
J_\alpha(z)=J_{\alpha\alpha}(z)
\end{align}

These generators can be split into $W_N$ and $\mathfrak{F}_a$ generators by the replacement
\begin{equation}
J_\alpha(z)\rightarrow J_\alpha(z)+j(z),
\end{equation}
where $j(z)$ is identified with the $U(1)$ current of $\mathfrak{F}$, while after the replacement $\sum J_\alpha=0$. We will however, for convenience consider directly the original $\widehat{\mathfrak{gl}(N)}_1$ currents.

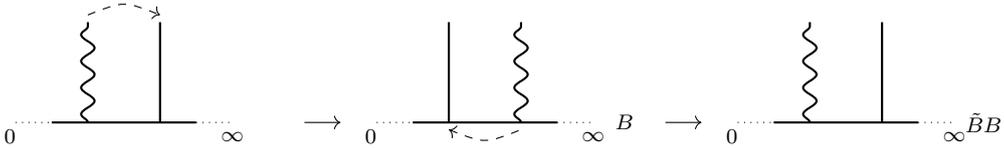
\begin{figure}[h!]
\begin{center}
\begin{tikzpicture}

\tikzmath{\scale=0.95;\radius=1;\length=1.4;\hlength=1;\ind=0.2;}


\begin{scope}[scale=\scale]
\begin{scope}
\draw[dotted] ( -1.5*\radius,0)--(-\radius,0) (\radius,0)--(1.5*\radius,0);
\draw[thick] (-\radius,0)--(\radius,0);
\draw[dashed,->] (-0.5*\radius,\length+0.1) ..controls (0,\length+0.3) and (0,\length+0.3).. (0.5*\radius,\length+0.1);
\draw[thick](0.5*\radius,0)--+(0,\length);
\draw[thick,wave](-0.5*\radius,0)--+(0,\length);
\draw[->](2.5,0)--(3,0);
\node[anchor=west] at(-1.8*\radius,-0.2) {\scriptsize $0$};
\node[anchor=east] at(1.8*\radius,-0.2) {\scriptsize $\infty$};
\end{scope}
\begin{scope}[xshift=5cm]
\draw[dashed,<-] (-0.5*\radius,-0.1) ..controls (0,-0.3) and (0,-0.3).. (0.5*\radius,-0.1);
\draw[dotted] ( -1.5*\radius,0)--(-\radius,0) (\radius,0)--(1.5*\radius,0);
\draw[thick] (-\radius,0)--(\radius,0);
\draw[thick,wave](0.5*\radius,0)--+(0,\length);
\draw[thick](-0.5*\radius,0)--+(0,\length);
\node[anchor=west] at(-1.8*\radius,-0.2) {\scriptsize $0$};
\node[anchor=east] at(1.8*\radius,-0.2) {\scriptsize $\infty$};
\node[anchor=east] at(2.2*\radius,0) {\scriptsize $B$};
\draw[->](2.5,0)--(3,0);
\end{scope}
\begin{scope}[xshift=10cm]
\draw[dotted] ( -1.5*\radius,0)--(-\radius,0) (\radius,0)--(1.5*\radius,0);
\draw[thick] (-\radius,0)--(\radius,0);
\draw[thick](0.5*\radius,0)--+(0,\length);
\draw[thick,wave](-0.5*\radius,0)--+(0,\length);
\node[anchor=west] at(-1.8*\radius,-0.2) {\scriptsize $0$};
\node[anchor=east] at(1.8*\radius,-0.2) {\scriptsize $\infty$};
\node[anchor=east] at(2.3*\radius,0) {\scriptsize $\tilde{B}B$};
\end{scope}
\end{scope}
\end{tikzpicture}
\end{center}
\caption{Braiding of a fermion with a vertex operator. The wavy line represents the insertion of a free fermion operator, while the solid line represents the insertion of a vertex.}
\label{Fig:FermionBraiding}
\end{figure}

As a consequence of what we just said, the fermionic Hilbert space $\mathcal{H}$ can be decomposed in sectors with definite $\widehat{\mathfrak{gl}(N)}_1$ charge given by a vector $\bs n\in\mathbb{Z}^N$:
\begin{equation}\label{eq:FermiHilbert}
\mathcal{H}=\bigoplus_{\bs n\in\mathbb{Z}^N}\mathcal{H}_{\bs n}.
\end{equation}
From the free fermions we can also define vertex operators in an axiomatic way by their braiding relations involving free fermions, i.e. as intertwiners (for more details, see \cite{Gavrylenko:2016moe}): if one analytically extends a matrix element involving $\psi(z)$ along a contour $\gamma$ that interchanges its time-ordering with a vertex operator $V_{\bs\theta}$ going counterclockwise above the insertion of the vertex operator, then
\begin{align}\label{eq:FermionVertexBraiding1}
\bar{\psi}(\gamma\cdot z)V_{\bs\theta}(0)=V_{\bs\theta}(0)B^{-1}\bar{\psi}(z), && \psi(\gamma\cdot z)V_{\bs\theta}(0) =V_{\bs\theta}(0)\psi(z)B.
\end{align}
Although our discussion will be fully general, the explicit form of $B$ is known, for $SL_N$, only for the specific semi-degenerate case,
that we will discuss in detail in Section \ref{sec:VerlindeSol}.
Let us denote by $\tilde{B}$ the braiding matrix defined by
\begin{align}
V_{\bs\theta}(0)\bar{\psi}(\tilde{\gamma}\cdot z)=\tilde{B}^{-1}\bar{\psi}(z)V_{\bs\theta}(0), && V_{\bs\theta}(0)\psi(\tilde{\gamma}\cdot z) =\psi(z)\tilde{B}V_{\bs\theta}(0),
\end{align}
where $\tilde{\gamma}$ follows the same orientation as $\gamma$, but goes below the insertion of the vertex operator: see the second and third step in Figure \ref{Fig:PunctureMonodromy}. Then we can compute the monodromies around any punctures by iterating these two moves, noting that $\tilde{\gamma}\circ\gamma$ represents a noncontractible contour around the point of insertion of the vertex:
\begin{equation}
\begin{split}
\langle\bs\sigma|\dots V_{\bs\theta}(z_k)\psi(z)\dots|\bs\sigma'\rangle&\rightarrow \langle\bs\sigma|\dots \psi(z)V_{\bs\theta}(z_k)\dots|\bs\sigma'\rangle\\
& = \langle\bs\sigma|\dots V_{\bs\theta}(z_k)\psi(z)\dots|\bs\sigma'\rangle B_k \\
&\rightarrow \langle\bs\sigma|\dots V_{\bs\theta}(0)\psi(z)\dots|\bs\sigma'\rangle \tilde{B}_kB_k.
\end{split}
\end{equation}
The monodromy as composition of braiding operation is represented pictorially in Figure \ref{Fig:FermionBraiding}.
To be able to compute all the monodromies, we also need a further ingredient: when the fermion is inserted near zero, its monodromy is diagonal, and given by~\footnote{In our notations $e^{2\pi i{\bs a}}={\rm diag}(e^{2\pi i a_i},\ldots, e^{2\pi i a_N})$}
\begin{equation}
\psi(\gamma_0\cdot z)|\bs a\rangle=\psi(z)|\bs a\rangle e^{2\pi i\bs a}.
\end{equation}
In fact, this is not only true for the primary state $|\bs a\rangle$ but also for all descendants
\begin{equation}
|\textbf{M},\bs a\rangle\equiv \psi_{\alpha_1,-p_1}\ldots \psi_{\alpha_l,-p_l}\bar\psi_{\beta_1,-q_1}\ldots \bar\psi_{\beta_l,-q_l}|\bs a\rangle,
\end{equation}
labeled by the coloured Maya diagram
\begin{equation}
\textbf{M}=\{((\alpha_1,-p_1),\ldots,(\alpha_l,-p_l)), ((\beta_1,-q_1),\ldots,(\beta_l,-q_l))\}.
\end{equation}
Analogous statements hold if the fermion is inserted instead near infinity. These last points follows from the solution of the problem on the three-punctured sphere: by repeated insertions of the identity
\begin{equation}
\langle\bs\sigma|\dots\psi(z)V_{\bs\theta}(z_k)\dots|\bs\sigma'\rangle=\sum_{\textbf{M},\textbf{M}'}\langle\bs\sigma|\dots|\textbf{M},\bs a\rangle\langle\textbf{M},\bs a| \psi(z)V_{\bs\theta}(z_k)|\textbf{M}',\bs a'\rangle\langle\textbf{M}',\bs a'|\dots|\bs\sigma'\rangle
\end{equation}
we can reduce the problem of computing monodromies around arbitrary punctures to a repeated use of the rules described above.

\begin{figure}[h!]
\begin{center}
\begin{tikzpicture}
\tikzmath{\scale=0.95;\radius=1;\length=1.4;\hlength=1;\ind=0.2;}


\begin{scope}[scale=\scale]
\begin{scope}
\draw[dotted] ( -1.5*\radius,0)--(-\radius,0) (\radius,0)--(1.5*\radius,0);
\draw[thick] (-\radius,0)--(\radius,0);
\draw[thick](0.5*\radius,0)--+(0,\length);
\draw[thick,wave](-0.5*\radius,0)--+(0,\length);
\draw[->](2.5,0)--(3,0);
\node[anchor=west] at(-1.8*\radius,-0.2) {\scriptsize $0$};
\node[anchor=east] at(1.8*\radius,-0.2) {\scriptsize $\infty$};
\end{scope}


\begin{scope}[xshift=5cm]
\draw[dotted] ( -1.5*\radius,0)--(-\radius,0) (\radius,0)--(1.5*\radius,0);
\draw[thick] (-\radius,0)--(\radius,0);
\draw[thick](0,0.5*\length)--+(\length*0.7,\length*0.7);
\draw[thick] (0,0)--+(0,0.5*\length) ;
\draw[thick,wave](0,0.5*\length)--+(-\length*0.7,\length*0.7);
\node[anchor=west] at(-1.8*\radius,-0.2) {\scriptsize $0$};
\node[anchor=east] at(1.8*\radius,-0.2) {\scriptsize $\infty$};
\node[anchor=east] at(2.2*\radius,0) {\scriptsize $F$};
\draw[->](2.5,0)--(3,0);
\end{scope}


\begin{scope}[xshift=10cm]
\draw[dotted] ( -1.5*\radius,0)--(-\radius,0) (\radius,0)--(1.5*\radius,0);
\draw[thick] (-\radius,0)--(\radius,0);
\draw[thick,wave](0,0.5*\length)--+(\length*0.7,\length*0.7);
\draw[thick] (0,0)--+(0,0.5*\length) ;
\draw[thick](0,0.5*\length)--+(-\length*0.7,\length*0.7);
\node[anchor=west] at(-1.8*\radius,-0.2) {\scriptsize $0$};
\node[anchor=east] at(1.8*\radius,-0.2) {\scriptsize $\infty$};
\node[anchor=east] at(2.7*\radius,0) {\scriptsize $e^{i\pi\bs\theta}F$};
\end{scope}


\begin{scope}[xshift=5cm,yshift=-3cm]
\draw[dotted] ( -1.5*\radius,0)--(-\radius,0) (\radius,0)--(1.5*\radius,0);
\draw[thick] (-\radius,0)--(\radius,0);
\draw[thick,wave](0.5*\radius,0)--+(0,\length);
\draw[thick](-0.5*\radius,0)--+(0,\length);
\draw[->](-2.5,0)--(-2,0);
\node[anchor=west] at(-1.8*\radius,-0.2) {\scriptsize $0$};
\node[anchor=east] at(1.8*\radius,-0.2) {\scriptsize $\infty$};
\node[anchor=east] at(3.4*\radius,0) {\scriptsize $\tilde{F}^{-1}e^{i\pi\bs\theta}F$};
\end{scope}

\end{scope}
\end{tikzpicture}
\end{center}
\caption{Four-point braiding as composition of three-point braiding and fusion.}
\label{Fig:FermionBraiding1}
\end{figure}
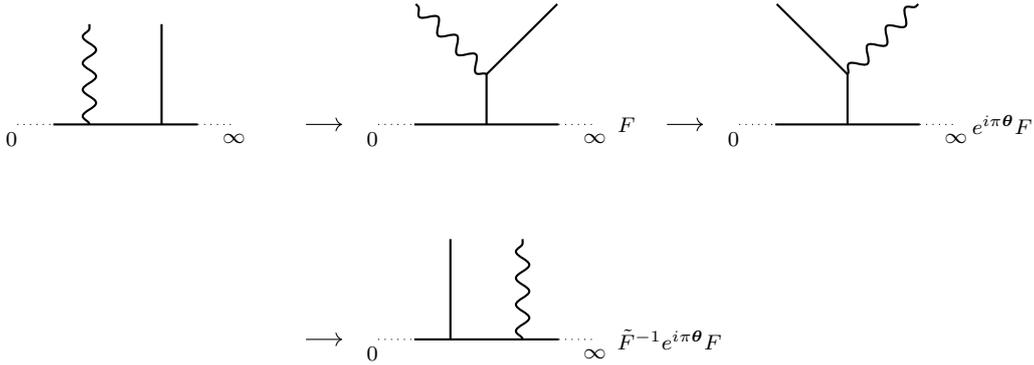

Finally, as shown in Figure \ref{Fig:FermionBraiding1}, let us note that the braiding matrix $B$ can be explicitly written in terms of the fusion matrix of the fermions with the vertex operators as
\begin{equation}
B_{\bs\theta}=F \left[ \begin{array}{cc} 1 & \psi \\ 0 & \infty 
\end{array} \right]^{-1}e^{i\pi\bs\theta_1}F \left[ \begin{array}{cc} \psi & 1 \\ 0 & \infty
\end{array} \right] \equiv\tilde{F}^{-1}e^{i\pi\bs\theta_1} F
\end{equation}
by decomposing the four-point braiding move into two fusion and one three-point braiding moves\footnote{We used here the standard notation of \cite{moore1989} for the fusion matrix.}. From this it is clear that the parameters $\bs\theta$ characterizing the vertex operators are the monodromy exponents of the linear system, since the monodromy around the vertex insertion has the form
\begin{equation}
M=\tilde{B}B=F^{-1}e^{2\pi i\bs\theta}F,
\end{equation}
so that by choosing different $\bs\theta$'s for the vertex operators we can realize monodromies in arbitrary conjugacy classes. Further note that an explicit form of $\tilde{B}, B$ is not actually needed to arrive to this conclusion: 
we will obtain in Section \ref{sec:VerlindeSol} the explicit form of the braiding matrix for the semi-degenerate case, which is given by equation \eqref{eq:BraidSemiDeg}.

\section{Kernel and tau function from free fermions}\label{sec:FreeFermSol}

We now show that the kernel \eqref{eq:KernelDef} has the following expression in terms of free fermion conformal blocks:
\begin{equation}\label{eq:RHKernel}
K(z',z)=Y^{-1}(z')\Xi(z-z', \textbf{Q})Y(z)=\frac{\langle V_{\bs\theta_1}(z_1)\dots V_{\bs\theta_n}(z_n)\bar{\psi}(z')\otimes\psi(z)\rangle}{\langle V_{\bs\theta_1}(z_1)\dots V_{\bs\theta_n}(z_n)\rangle},
\end{equation}
where
\begin{equation}
\Xi(z-z',\textbf{Q})=\diag\left(x(\sigma\tau+\rho-Q_1,z),\dots,x(\sigma\tau+\rho-Q_n,z) \right],
\end{equation}
$x$ being the Lam\'e function defined in Appendix \ref{sec:Theta}. The notation $\langle\dots\rangle$ stands for
\begin{equation}
\langle\mathcal{O}\rangle=\tr_{\mathcal{H}}\left(q^{L_0}(-)^Fe^{2\pi i\bs\eta\cdot\bs J_0} \mathcal{O}\right),
\end{equation}
where $\mathcal{H}$ is our free fermionic Hilbert space \eqref{eq:FermiHilbert}, $J_0^i$ are the $\widehat{\mathfrak{gl}(N)}_1$ Cartan charges and $\eta_i$ their fugacities. The insertion of $(-)^F$ shifts the periodicity condition of our fermions around the B-cycle of the torus, and will be relevant in the computation of the B-cycle monodromy. As discussed in Section \ref{sec:LinSys}, we included the $U(1)$ charge and fugacity in the definition of $\Xi$, that we denoted by
\begin{align}
\sigma=\frac1N\sum_{i=1}^N\sigma_i, && \rho=\frac1N\sum_{i=1}^N\eta_i.
\end{align}
It will be also useful to introduce $\mathfrak{sl}_n$ projections of the charge vectors
\begin{align}
\tilde\sigma_i=\sigma_i-\sigma,&&\tilde\eta_i=\eta_i-\rho.
\end{align}

The motivation behind the matrix $\Xi$ is the following: it gives the LHS of the equation a simple pole, that in the RHS is due to the OPE of the free fermions, while also producing the $U(1)$ part of the monodromies, absent in $Y$ but present by construction in the CFT. Further and most importantly, it cancels both the twists of the solution $Y$, so that the kernel $K$ has monodromies acting from both left and right as in equation \eqref{eq:KernelMonodromy}.
Our goal will be to show that the vertex operators can be defined in such a way that the RHS has given monodromies acting in exactly such a way with prescribed conjugacy class, which together with the identical singular behavior around $z,z'\sim z_k$, $z\sim z'$ coming from the OPE of the free fermions with the vertex operators shows that the two objects coincide.

\begin{figure}[h!]
\begin{center}
\begin{tikzpicture}

\tikzmath{\scale=0.95;\radius=0.9;\length=1.5;\hlength=0.8;\ind=0.2;}

\begin{scope}[scale=\scale]

\begin{scope}
\draw[rounded corners=\radius cm*\scale, thick] (0,\radius)--(-\radius-\hlength,\radius)--(-\radius-\hlength,-\radius)--
(\radius+\hlength,-\radius)--(\radius+\hlength,\radius)--(0,\radius);
\draw[thick,wave](\hlength*0.9,\radius)--+(0,\length);
\draw[thick,wave](\hlength*0.3,\radius)--+(0,\length);
\draw[thick](-\hlength*0.3,\radius)--+(0,\length);
\draw[thick](-\hlength*0.9,\radius)--+(0,\length);
\draw[dashed,<->] (\hlength*0.4,\length+\radius+0.1) ..controls (0,\length+\radius+0.2) and (0,\length+\radius+0.2).. (-\hlength*0.4,\length+\radius+0.1);
\draw[->](1.8,0)--(2.2,0);
\end{scope}

\begin{scope}[xshift=4cm]
\draw[rounded corners=\radius cm*\scale, thick] (0,\radius)--(-\radius-\hlength,\radius)--(-\radius-\hlength,-\radius)--
(\radius+\hlength,-\radius)--(\radius+\hlength,\radius)--(0,\radius);
\draw[thick,wave](\hlength*0.9,\radius)--+(0,\length);
\draw[thick](\hlength*0.3,\radius)--+(0,\length);
\draw[thick,wave](-\hlength*0.3,\radius)--+(0,\length);
\draw[thick](-\hlength*0.9,\radius)--+(0,\length);
\draw[dashed,<->] (-\hlength*0.2,\length+\radius+0.1) ..controls (-\hlength*0.6,\length+\radius+0.2) and (-\hlength*0.6,\length+\radius+0.2).. (-\hlength*1,\length+\radius+0.1);
\draw[->](1.8,0)--(2.2,0);
\end{scope}

\begin{scope}[xshift=8cm]
\draw[rounded corners=\radius cm*\scale, thick] (0,\radius)--(-\radius-\hlength,\radius)--(-\radius-\hlength,-\radius)--
(\radius+\hlength,-\radius)--(\radius+\hlength,\radius)--(0,\radius);
\draw[thick,wave](\hlength*0.9,\radius)--+(0,\length);
\draw[thick](\hlength*0.3,\radius)--+(0,\length);
\draw[thick](-\hlength*0.3,\radius)--+(0,\length);
\draw[thick,wave](-\hlength*0.9,\radius)--+(0,\length);
\draw[dashed,<->] (-\hlength,\radius-0.1) ..controls (-\hlength*0.5,\radius-0.3) and (-\hlength*0.5,\radius-0.3).. (0,\radius-0.1);
\draw[->](1.8,0)--(2.2,0);
\end{scope}

\begin{scope}[xshift=12cm]
\draw[rounded corners=\radius cm*\scale, thick] (0,\radius)--(-\radius-\hlength,\radius)--(-\radius-\hlength,-\radius)--
(\radius+\hlength,-\radius)--(\radius+\hlength,\radius)--(0,\radius);
\draw[thick,wave](\hlength*0.9,\radius)--+(0,\length);
\draw[thick](\hlength*0.3,\radius)--+(0,\length);
\draw[thick,wave](-\hlength*0.3,\radius)--+(0,\length);
\draw[thick](-\hlength*0.9,\radius)--+(0,\length);
\draw[dashed,<->] (-\hlength*0.5,\length+\radius+0.1) ..controls (0,\length+\radius+0.2) and (0,\length+\radius+0.2).. (\hlength*0.5,\length+\radius+0.1);
\end{scope}
\end{scope}
\end{tikzpicture}

\begin{tikzpicture}[thick]








\draw(0,0) circle[x radius=5.7cm,y radius=1.5cm];

\begin{scope}[xscale=6,yscale=3,yshift=1.4cm]
\draw(-0.7,-1.4) to[out= -30,in=210] (0.7,-1.4);
\draw(-0.55,-1.469) to[out= 30,in=-210] (0.55,-1.469);
\end{scope}

\draw[olive,-Stealth](1,0.7) to[out=143,in=10] (-1.7,1);

\draw[olive,-Stealth](-1.9,1) to[out=180,in=40] (-3.95,0.6);

\draw[olive,-Stealth](-4,0.5) to[out=-80,in=190] (-2.1,0.6);

\draw[olive,-Stealth](-2.1,0.6) to[out=30,in=170] (0.95,0.7);

\begin{scope}[xshift=-1.8cm,yshift=1cm,olive,dashed]
\draw[fill=white,white](0,0) circle[x radius=0.08,y radius=0.04];
\draw(0,0) circle[x radius=0.08,y radius=0.04];
\fill[white](0.08,-0)--(0.08,1.5)--(-0.08,1.5)--(-0.08,0);
\draw(0,1.5)circle[x radius=0.08,y radius=0.04];
\draw(0.08,0)--(0.08,1.5) (-0.08,0)--(-0.08,1.5);
\end{scope}

\begin{scope}[xshift=-4cm,yshift=0.5cm,olive,dashed]
  \draw[fill=white,white](0,0) circle[x radius=0.08,y radius=0.04];
\draw(0,0) circle[x radius=0.08,y radius=0.04];
\fill[white](0.08,0)--(0.08,1.5)--(-0.08,1.5)--(-0.08,0);
\draw(0,1.5)circle[x radius=0.08,y radius=0.04];
\draw(0.08,0)--(0.08,1.5) (-0.08,0)--(-0.08,1.5);
\end{scope}

\begin{scope}[xshift=-2.05cm,yshift=0.6cm,olive,dashed]
  \draw[fill=white,white](0,0) circle[x radius=0.08,y radius=0.04];
\draw(0,0) circle[x radius=0.08,y radius=0.04];
\fill[white](0.08,0)--(0.08,1.5)--(-0.08,1.5)--(-0.08,0);
\draw(0,1.5)circle[x radius=0.08,y radius=0.04];
\draw(0.08,0)--(0.08,1.5) (-0.08,0)--(-0.08,1.5);
\end{scope}

\begin{scope}[xshift=-3 cm,yshift=0.4 cm]
  \fill[white](-0.6,0) to [out=35,in=-90](-0.2,1.5) to (0.2,1.5) to[out=-90,in=145] (0.6,0.2)--cycle;

\draw(-0.6,0) to [out=35,in=-90](-0.2,1.5)
(0.6,0.2) to [out=145,in=-90](0.2,1.5);
\draw(0,1.5)circle[x radius=0.2,y radius=0.1];
\end{scope}

\begin{scope}[xshift=-1 cm,yshift=0.7 cm]
  \fill[white] (-0.6,0) to [out=35,in=-90](-0.2,1.5) to (0.2,1.5) to[out=-90,in=145] (0.6,0)--cycle;

\draw(-0.6,0) to [out=35,in=-90](-0.2,1.5)
(0.6,0) to [out=145,in=-90](0.2,1.5);
\draw(0,1.5)circle[x radius=0.2,y radius=0.1];
\end{scope}

\begin{scope}[xshift=1cm,yshift=0.7cm]
    \draw[fill=white,white](0,0) circle[x radius=0.08,y radius=0.04];
\draw(0,0) circle[x radius=0.08,y radius=0.04];
\fill[white](0.08,0)--(0.08,1.5)--(-0.08,1.5)--(-0.08,0);
\draw(0,1.5)circle[x radius=0.08,y radius=0.04];
\draw(0.08,0)--(0.08,1.5) (-0.08,0)--(-0.08,1.5);
\end{scope}

\begin{scope}[xshift=2.8cm,yshift=0.4cm]
\draw(0,0) circle[x radius=0.08,y radius=0.04];
\fill[white](0.08,0)--(0.08,1.5)--(-0.08,1.5)--(-0.08,0);
\draw(0,1.5)circle[x radius=0.08,y radius=0.04];
\draw(0.08,0)--(0.08,1.5) (-0.08,0)--(-0.08,1.5);
\end{scope}

\end{tikzpicture}

\end{center}
\caption{Monodromy of a fermion around a puncture through braiding on the two-punctured torus. On the upper side, the steps that compose the monodromy operation are represented in terms of conformal block diagrams. On the lower side, the meaning of the conformal block diagram is drawn on the torus: the thin cylinders represent the fermions, while the larger tubes represent the vertex operators. The intermediate steps are drawn in olive green.}
\label{Fig:PunctureMonodromy}
\end{figure}
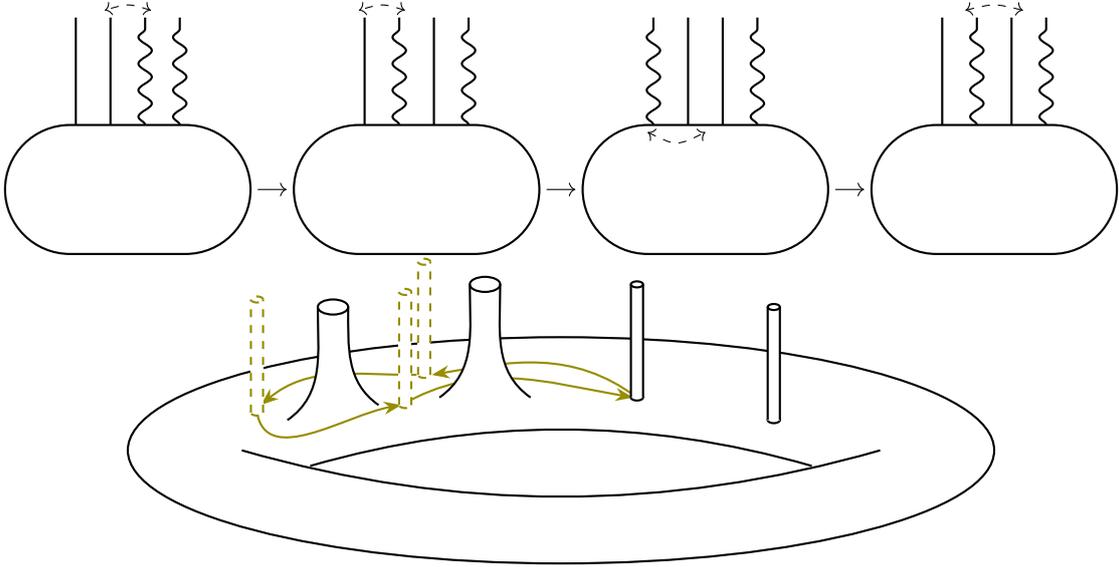

In this section we compute the monodromies following the method explained in Section \ref{sec:FreeFerm}: the vertex operators are defined through their action on free fermions, so it is possible to realize a monodromy with prescribed conjugacy class at every puncture. Operationally, if one wants to compute the monodromy around the cycle $\gamma_n$, for example, the operation is the following (we are summing over repeated indices):
\begin{equation}
\begin{split}
\langle V_{\bs\theta_1}(z_1)\dots V_{\bs\theta_n}(z_n)\bar{\psi}_i(z')\psi_j(z)\rangle & \rightarrow-\langle V_{\bs\theta_1}(z_1)\dots V_{\bs\theta_n}(z_n)\psi_j(z)\bar{\psi}_i(z')\rangle \\
& \rightarrow -\langle V_{\bs\theta_1}(z_1)\dots\psi_k(z) V_{\bs\theta_n}(z_n)\bar{\psi}_i(z')\rangle (B_n){^k}_{j} \\
& \rightarrow-\langle V_{\bs\theta_1}(z_1)\dots V_{\bs\theta_n}(z_n)\psi_k(z)\bar{\psi}_i(z')\rangle (\tilde{B}_nB_n){^k}_{j} \\
& \rightarrow\langle V_{\bs\theta_1}(z_1)\dots V_{\bs\theta_n}(z_n)\bar{\psi}_i(z')\psi_k(z)\rangle (\tilde{B}_nB_n){^k}_{j},
\end{split}
\end{equation}
so that the monodromy around $z_n$ is
\begin{equation}
M_n=\tilde{B}_nB_n=F_n^{-1}e^{2\pi i\bs\theta_n}F_n \sim e^{2\pi i\bs\theta_n} .
\end{equation}
Following the same idea, one can compute the monodromy around an arbitrary puncture $z_\alpha$: one perform a braiding around every puncture from $z_n$ to $z_{\alpha+1}$, then twice around $z_\alpha$, then again around $z_{\alpha}$ to $z_n$ in the opposite direction as before. The operation is represented graphically in Figure \ref{Fig:PunctureMonodromy} for the puncture $z_{1}$ in the two-punctured torus.
The result is that the monodromy around an arbitrary puncture $z_\alpha$ is given by
\begin{equation}
\begin{split}
M_\alpha & = B_n^{-1}\dots B_{\alpha+1}^{-1}\tilde{B}_\alpha B_\alpha B_{\alpha+1}\dots B_n\\
& =(F_\alpha B_{\alpha+1}\dots B_n)^{-1}e^{2\pi i\bs\theta_\alpha}(F_{\alpha}B_{\alpha+1}\dots B_n)\sim e^{2\pi i\bs\theta_\alpha}.
\end{split}
\end{equation}
The monodromy around the A-cycle is fixed by our choice of gluing: it is given by
\begin{equation}
M_A=e^{2\pi i \bs a}.
\end{equation}

Finally, the monodromy around the B-cycle can be computed in the following way. First we go once around every $z_k$:
\begin{equation}
\begin{split}
\langle V_{\bs\theta_1}(z_1)\dots V_{\bs\theta_n}(z_n)& \bar{\psi}_i(z')\psi_j(z)\rangle\rightarrow-\langle V_{\bs\theta_1}(z_1)\dots V_{\bs\theta_n}(z_n)\psi_j(z)\bar{\psi}_i(z')\rangle \\
& \rightarrow\dots\rightarrow-\langle\psi_k(z) V_{\bs\theta_1}(z_1)\dots V_{\bs\theta_n}(z_n)\bar{\psi}_i(z')\rangle(B_1\dots B_n){^k}_j.
\end{split}
\end{equation}
Now, to go around the B-cycle we have to bring the fermion back to the original position without crossing again the other operators. This is done by using the cyclicity of the trace, but in fact in doing so we also have to take into account the insertion of $(-)^Fe^{2\pi i\bs\eta\cdot\bs J_0}$ in the trace:
\begin{equation}
\begin{split}
&-\langle\psi_k(z) V_{\bs\theta_1}(z_1)\dots V_{\bs\theta_n}(z_n)\bar{\psi}_i(z')\rangle(B_1\dots B_n){^k}_j\\
& =-\tr_{\mathcal{H}}\left(q^{L_0}(-)^Fe^{2\pi i\bs\eta\cdot\bs J_0}\psi_k(z) V_{\bs\theta_1}(z_1)\dots V_{\bs\theta_n}(z_n)\bar{\psi}_i(z')\right)(B_1\dots B_n){^k}_j \\
& \rightarrow \tr_{\mathcal{H}}\left(\psi_k(z)q^{L_0}(-)^Fe^{2\pi i\bs\eta\cdot\bs J_0} V_{\bs\theta_1}(z_1)\dots V_{\bs\theta_n}(z_n)\bar{\psi}_i(z')\right) (e^{2\pi i\bs\eta}B_1\dots B_n){^k}_j \\
& = \langle V_{\bs\theta_1}(z_1)\dots V_{\bs\theta_n}(z_n)\bar{\psi}_i(z')\psi_k(z)\rangle\cdot e^{2\pi i\rho} (e^{2\pi i\tilde{\bs\eta}}B_1\dots B_n){^k}_j
\end{split}
\end{equation}
so that
\begin{equation}
M_B=e^{2\pi i\rho} e^{2\pi i\tilde{\bs\eta}}B_1\dots B_n.
\end{equation}
The two sides of equation \eqref{eq:RHKernel} have prescribed monodromies and singular behavior, and so they coincide. To compute the tau function we have to expand the trace of equation \eqref{eq:RHKernel} for $z\sim z'$. 

By expanding the LHS, we get a term involving the Lax matrix
\begin{equation}
Y(z+t/2)Y^{-1}(z-t/2)=\left(\mathbb{I}+tL(z)+\frac{t^2}{2}L^2(z) \right),
\end{equation}
and two terms from the expansion of the matrix $\Xi$:
\begin{equation}
\frac{\theta_1'(0)}{\theta_1(t)}=\frac{1}{t}-\frac{t}{6}\frac{\theta_1'''}{\theta_1'}+O(t^3) ,
\end{equation}
\begin{equation}
\frac{\theta_1(t-\tilde{Q}_i)}{\theta_1(-\tilde{Q}_i)}=1+t\frac{\theta_1'(-\tilde{Q}_i)}{\theta_1(-\tilde{Q}_i)}+\frac{t^2}{2}\frac{\theta_1''(-\tilde{Q}_i)}{\theta_1(-\tilde{Q}_i)}.
\end{equation}
Here we introduced
\begin{equation}
\tilde{Q}_i=Q_i-\sigma\tau-\rho.
\end{equation}
On the RHS, the expansion consists of the OPE for the fermions, yielding
\begin{equation}\label{eq:OPEFermions}
\tr\psi(z+t/2)\otimes\bar{\psi}(z-t/2) =\frac{N}{t}+Nj(z)+\frac{t}{2}T(z)+O(t^2).
\end{equation}

The $O(t)$ term relates the expectation value of the energy-momentum tensor to the trace squared of the Lax matrix:
\eq{\label{eq:EMInsertion}
\frac{\langle T(z) V_1\ldots V_n\rangle}{\langle V_1\ldots  V_n\rangle}=\frac12\tr  L^2(z)+\tr  L(z)\frac{\theta_1'(\tilde{\bs Q})}{\theta_1(\tilde{\bs Q})}+
\frac12\tr\frac{\theta_1''(\tilde{\bs Q})}{\theta_1(\tilde{\bs Q})}-\frac{N}{6}\frac{\theta_1'''(0)}{\theta_1'(0)}\\
\equiv\frac12 \tr L^2(z)+t(z)
}
We see that, as in \cite{Bonelli:2019boe}, in the genus one case there is a correction to the relation that one has in genus zero \cite{Iorgov:2014vla,Gavrylenko:2016moe,Gavrylenko:2018ckn}, encoded in $t(z)$.

We wish now to determine the expression for the tau function  by computing contour integrals of \eqref{eq:EMInsertion} and comparing with \eqref{LaxHamilt2} and \eqref{eq:TauHamilt}. From \eqref{eq:EMInsertion} we see that we can split the tau function in two parts:
\begin{equation}
\Tau=\Tau_0\Tau_1,
\end{equation}
which are defined by the following equations:
\begin{align}
\partial_{z_k}\log\Tau_0=\oint_{\gamma_k}\frac{dz}{2\pi i}\langle T(z) V_1\ldots V_n\rangle, && 2\pi i \partial_\tau\log\Tau_0=\oint_Adz\frac{1}{2}\langle T(z) V_1\ldots V_n\rangle,
\end{align}
\begin{align}
\partial_{z_k}\log\Tau_1=-\oint_{\gamma_k}\frac{dz}{2\pi i}t(z), && 2\pi i \partial_\tau\log\Tau_1=-\oint_Adzt(z).
\end{align}
The first term would be there also in the genus zero case, while the second term is a new feature appearing in higher genus. $\Tau_0$ is computed by applying the Virasoro Ward identity:
\begin{equation}
\langle T(z) V_1\ldots V_n\rangle=\langle T\rangle+\sum_{k=1}^n E_1(z-z_k)\partial_k\log\langle V_1\ldots V_n\rangle +\sum_{k=1}^n\bs\theta_k^2 E_2(z-a_k),
\end{equation}
yielding
\begin{equation}
\Tau_0=\langle V_1\dots V_n\rangle.
\end{equation}

We now turn to computing the contour integrals of $t(z)$: since we have
\begin{align}
\sum_i S_{ii}^{(k)}=0, && \int_0^1dz\frac{\theta_1'(z-z_k)}{\theta_1(z-z_k)}=\pi i
\end{align}
when $z_k$ lies in the fundamental domain. Then, the only contribution to the $\tau$-derivative of $\Tau_1$ will be
\eq{
-2\pi i\partial_\tau\mc\log\mc T_1=\tr \bs p\frac{\theta_1'(\tilde{\bs Q})}{\theta_1(\tilde{\bs Q})}+\frac12\tr\frac{\theta''_1(\tilde{\bs Q})}{\theta_1(\tilde{\bs Q})}
-\frac{N}6\frac{\theta_1'''(0)}{\theta_1'(0)}=\\
=2\pi i\tr \partial_\tau\tilde{\bs Q} \frac{\theta_1'(\tilde{\bs Q})}{\theta_1(\tilde{\bs Q})}+2\pi i
\tr \frac{\partial_\tau\theta_1(\tilde{\bs Q})}{\theta_1(\tilde{\bs Q})}-
2\pi i\frac N3\frac{\partial_\tau\theta'(0)}{\theta_1'(0)}=
2\pi i\partial_\tau\left(\tr \log\theta_1(\tilde{\bs Q})-N\log\eta(\tau)\right)
}
Therefore
\eq{\label{eq:IntConstant}
\mc T_1=f(\{z_k\})\frac{\eta(\tau)^N}{\prod_i\theta_1(\tilde{Q}_i(\{z_k\},\tau))}=\frac{f(\{z_k\})}{Z_{twist}(\tilde{\bs Q}(\{z_k\},\tau))} ,
}
where $f(\{z_k\})$ is an arbitrary function of the punctures' positions, left undetermined by the integration. In fact, let us show that $f(\{z_k\})=1$: computing the residues of $t(z)$ yields
\eq{
-\partial_{z_k}\log\mc T_1=\sum_i S_{ii}^{(k)}\frac{\theta_1'(\tilde{Q}_i)}{\theta_1(\tilde{Q}_i)}.
}
At first sight, the RHS doesn't look like a total $z_k$-derivative. However, let us consider the $\bs p$-dependent part of the corresponding Hamiltonian:
\eq{
H_k=\frac12\res_{z_k} \tr \mc A(z)^2=\sum_i S_{ii}^{(k)} p_i + \ldots
}
from which it follows that $\partial_{z_k}Q_i=S_{ii}^{(k)}$. Therefore
\eq{
\partial_{z_k}\log\mc T_1=\sum_i\frac{\theta_1'(\tilde{Q}_i)}{\theta_1(\tilde{Q}_i)}\partial_{z_k}\tilde{Q}_i=\partial_{z_k}\log\prod_i\theta_1(\tilde{Q}_i)
}
Therefore in \eqref{eq:IntConstant} $f(\{z_k\})=const$, and we can put without loss of generality $f(\{z_k\})=1$, as promised. The isomonodromic tau function is
\begin{equation}\label{eq:TauFunctionCFT}
\Tau(\{z_k\},\tau)=\frac{1}{Z_{twist}(\tilde{\bs Q}(\tau))}\langle V_1(z_1) \dots V_n(z_n)\rangle.
\end{equation}
Let us remark that the CFT arguments used above are valid for general vertex insertions. However, in order to have explicit calculable expressions one needs to consider the insertion of (semi-) degenerate fields. In this case, the fermionic correlator is identified with the dual partition function of a circular quiver gauge theory with gauge group $U(N)^n $ and $n$ hypermultiplets in bifundamental representations of the gauge groups, as encoded in the conformal block diagram. Therefore, the above equality can be rewritten as
\begin{equation}\label{eq:TauFunctionGaugeU}
\Tau=\frac{Z^D(\tau,\{z_k\}|\{\bs a_k\},\{\bs\eta_k\},\{\bs\theta_k\})}{Z_{twist}(\tilde{\bs Q}(\tau,\{z_k\}))},
\end{equation}
where we made explicit the dependence on all the intermediate channel charges $\bs a_k$, $k=1,\dots n$, together with their duals entering in the Fourier transform $\bs\eta_k$, and set $\bs a\equiv\bs a_1$.

\section{Torus monodromies with Verlinde loop operators}\label{sec:VerlindeSol}
In this section we show an alternative proof of formulas \eqref{eq:RHKernel} and \eqref{eq:TauFunctionCFT} for the kernel and tau function respectively, using Verlinde loop operators acting on (semi-) degenerate representations of $W_N$ algebras, along the lines of \cite{Gavrylenko:2018ckn}. The necessary definitions about degenerate fields and $W_N$ algebras are collected in Appendix \ref{sec:WNApp}.

\subsection{General setup}

\begin{figure}[h!]
\begin{center}
\begin{tikzpicture}

\tikzmath{\scale=1;\radius=1;\length=1;\hlength=1.2;\ind=0.2;}

\begin{scope}[scale=\scale,xscale=2]

\draw[rounded corners=1cm*\scale, thick] (0,\radius)--(-\radius-\hlength,\radius)--(-\radius-\hlength,-\radius)--
(\radius+\hlength,-\radius)--(\radius+\hlength,\radius)--(0,\radius);

\draw[thick](0,\radius)--+(0,\length);
\draw[thick](0.5,\radius)--+(0,\length);
\draw[thick,wave](-0.5,\radius)--+(0,\length);

\draw[thick,wave](-1,\radius)--+(0,\length);

\node[rotate=-45,anchor=west] at(0.2,\radius) {\scriptsize $\bs\sigma^{1}$};
\node[rotate=-45,anchor=west] at(-0.3,\radius) {\scriptsize $\bs\sigma^{0}-\bs h_i+\bs h_j$};
\node[rotate=-45,anchor=west] at(-0.85,\radius) {\scriptsize $\bs\sigma^{0}-\bs h_i$};
\node[rotate=-45,anchor=west] at(-1.6,\radius) {\scriptsize $\bs\sigma^{0}$};

\node[rotate=45,anchor=west] at(-0.2,\length+\radius) {$\nu_1\bs\omega_1$};
\node[rotate=45,anchor=west] at(0.3,\length+\radius) {$\nu_2\bs\omega_1$};
\node[rotate=45,anchor=west] at(-0.7,\length+\radius) {$\bs\omega_{N-1}$};
\node[rotate=45,anchor=west] at(-1.2,\length+\radius) {$\bs\omega_{1}$};

\end{scope}

\end{tikzpicture}
\end{center}
\caption{Toric conformal blocks with $n=2$ semi-degenerate and 2 degenerate fields.}
\end{figure}
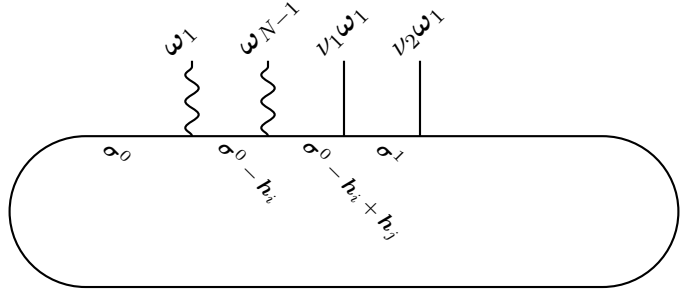

We wish to study the monodromy properties of the torus conformal block with insertions of two $W_N$ completely degenerate fields, $\phi$ and $\bar\phi$, and $n$ semi-degenerate W-primaries $V$:
\eq{
\label{eq:toric_block}
\Psi_{ij}(\bs\sigma^0;\bs\sigma^1,\ldots,\bs\sigma^{n-1}|z,z_0)=\\=
\tr_{\mathcal H_{\bs \sigma^0}}\left( q^{L_0}\phi_i(z)\bar\phi_j(z_0) V_{\nu_1}(z_1)\mc P_{\bs\sigma^1}V_{\nu_2}(z_2)\ldots V_{\nu_{n-1}}(z_{n-1}) 
\mc P_{\bs\sigma^{n-1}}V_{\nu_n}(z_n)\right).
}
In this formula the operators $V_{\nu_k}$ are semi-degenerate  W-primaries with W-charges given by $\bs\theta_k=\nu_k\bs\omega_1$, where $\bs\omega_1$ is the first fundamental weight of $A_{N-1}$. Operators $\phi_i$ and $\bar\phi_j$ are completely degenerate fields with W-charges given by $\bs\omega_1$ and $\bs\omega_{N-1}$, respectively. Indices $i$ and $j$ label fusion channels.

The normalization of $V_{\nu_k}$ is given by:
\eq{
\label{eq:normalization}
\langle \bs\sigma'|V_{\nu}|\bs\sigma\rangle\equiv\mc N^{+}(\bs\sigma',\nu\bs\omega_1,\bs\sigma),
}
where
\eq{
\label{eq:normalization}
\mc N^{\pm}(\bs\sigma',\nu\bs\omega_1,\bs\sigma)=\frac{\prod_{lj}G(1\mp\nu/N\pm\sigma_l\mp\sigma'_j)}{\prod_{k<m}G(1+\sigma_k-\sigma_m)G(1-\sigma'_k+\sigma'_m)}.
}
We also fix normalization of the completely degenerate field by\footnote{This parameterization differs from one in \cite{Gavrylenko:2018ckn} by the factor $e^{i\pi (1-N)(\bs\sigma,\bs h_i)}$.}
\eq{
  \langle\bs\sigma|\phi_i(1)|\bs\sigma-\bs h_i\rangle=e^{i\pi N(\bs\sigma,\bs h_i)}\mc N^-(\bs\sigma,\bs\omega_1,\bs\sigma-\bs h_i)\,.
  \label{eq:degenerate_norm}
}
As in equation \eqref{eq:ReducedDeg}, $\mc P_{\bs\sigma^k}$ is the projection operator onto the W-algebra representation with charge $\bs\sigma^k$, expressing the fact that the conformal block has fixed intermediate charges. 
It is useful to expand the trace of \eqref{eq:toric_block} as a sum of diagonal matrix elements:
\eq{
\Psi(\bs\sigma^0;\bs\sigma^1,\ldots,\bs\sigma^{n-1}|z,z_0)=
\sum_{\bs Y} q^{|\bs Y|} \Psi^{(\bs Y)}(\bs\sigma^0;\bs\sigma^1,\ldots,\bs\sigma^{n-1}|z,z_0)
}
where vector of Young diagrams $\bs Y$ labels W-algebra descendants, and we defined the matrix element between descendants
\eq{
\Psi^{(\bs Y)}
=\langle\bs\sigma^0,\bs Y|\phi(z)\otimes\bar\phi(z_0) V_{\nu_1}\ldots V_{\nu_n} |\bs\sigma^0,\bs Y\rangle\,.
}

We remind one of the the main results of \cite[Theorem 5.1]{Gavrylenko:2018ckn}: the Fourier transformation of $\Psi^{(\bs Y)}$ over all internal W-charges has number-valued (not operator-valued as generically happens) monodromies around $0,\infty$ and the insertion points $z_1,\dots, z_n$, as a function of
$z$ and $z_0$, independent from $\bs Y$. The Fourier transform is defined by
\eq{
\Psi^{(\bs Y)D}(\bs\sigma^0;\bs\sigma^1,\bs\eta^1,\ldots,\bs\sigma^{n-1},\bs\eta^{n-1}|z,z_0)
=\\=
\sum_{\bs w^i\in Q_{A_{N-1}}}e^{2\pi i\sum_{i=1}^{n-1} (\bs\eta^i,\bs w^i)}
\Psi^{(\bs Y)}(\bs\sigma^0;\bs\sigma^1+\bs w^1,\ldots,\bs\sigma^{n-1}+\bs w^{n-1}|z,z_0),
}
where $Q_{A_{N-1}}$ is the $\mathfrak{sl}_N$ root lattice. Moreover, for the case $\bs Y=\bs\emptyset$ the function $\Psi^{0;D}$ gives the solution of the $n+2$ point Fuchsian system on the sphere.
So using the results of \cite{Gavrylenko:2018ckn} we get automatically the following statement: the function 
$\Psi^{D}$, given by the formula\footnote{In all formulas letter ``$D$'' stands for ``dual''.}
\eq{
\Psi^{D}
=
\sum_{\bs w^i\in Q_{A_{N-1}}}\sum_{\bs Y}e^{2\pi i\sum_{i=1}^{n-1} (\bs\eta^i,\bs w^i)}
\Psi(\bs\sigma^0;\bs\sigma^1+\bs w^1,\ldots,\bs\sigma^{n-1}+\bs w^{n-1}|z,z_0)
}
has number-valued monodromies $M_k$ around all $z_k$, and also number-valued A-cycle monodromy $M_A=e^{2\pi i\bs\sigma^0}$,
since after taking trace we identify $A$-cycle with the loop around $0$ or $\infty$ on the initial sphere. The problem now is to find a linear combination of $\Psi^{D}$ that has number-valued 
monodromy around the B-cycle.

\subsection{B-cycle monodromy operator}

The main ingredient in the computation, as in the case of free fermions, is the braiding move exchanging two insertions in a four-point conformal block, as in Figure \ref{fig:fusion}, where we see how the braiding can be expressed in terms of the fusion matrix ${\sf B}$, given below in equation \eqref{eq:DegFusionKer}.

\begin{figure}[h!]
\begin{center}
\begin{tikzpicture}

\begin{scope}
\draw[thick](-1,0)--(1,0);
\draw[thick](0,0)--(0,1);
\draw[wave](0.5,0)--(0.5,1);

\node at(1,0.3) {$\bs\sigma$};
\node at(-1,0.3) {$\bs\sigma'$};
\node[rotate=-45,anchor=west] at(0.1,0){$\bs\sigma+\bs h_l$};

\node at(-0.1,1.15){$\nu \bs\omega_1$};
\node at(0.6,1.15){$\bs\omega_1$};
\end{scope}

\begin{scope}[xshift=1.1cm]
\node[anchor=west] at (0,0){$=\sum_j{\sf B}_{lj}(\bs\sigma',\nu,\bs\sigma)$};

\end{scope}

\begin{scope}[xshift=5.5cm]
\draw[thick](-1,0)--(1,0);
\draw[thick](0,0)--(0,1);
\draw[wave](-0.5,0)--(-0.5,1);

\node at(1,0.3) {$\bs\sigma$};
\node at(-1,0.3) {$\bs\sigma'$};
\node[rotate=-45,anchor=west] at(-0.4,0){$\bs\sigma'-\bs h_j$};

\node at(0.2,1.15){$\nu \bs\omega_1$};
\node at(-0.5,1.15){$\bs\omega_1$};
\end{scope}

\end{tikzpicture}
\end{center}
\caption{Fusion transformation of conformal blocks.}
\label{fig:fusion}
\end{figure}

\begin{figure}[h!]
\begin{center}
\begin{tikzpicture}

\tikzmath{\scale=0.95;\radius=1;\length=1.5;\hlength=1.2;\ind=0.2;}

\begin{scope}[scale=\scale]
\begin{scope}

\draw[rounded corners=1cm*\scale, thick] (0,\radius)--(-\radius-\hlength,\radius)--(-\radius-\hlength,-\radius)--
(\radius+\hlength,-\radius)--(\radius+\hlength,\radius)--(0,\radius);

\draw[rounded corners=1.2cm*\scale,dashed,->] (-1.1,\radius+\ind)--(-\radius-\hlength-\ind,\radius+\ind)--(-\radius-\hlength-\ind,-\radius-\ind)--
(\radius+\hlength+\ind,-\radius-\ind)--(\radius+\hlength+\ind,\radius+\ind)--(1.1,\radius+\ind);

\draw[thick](0,\radius)--+(0,\length);
\draw[thick](0.5,\radius)--+(0,\length);
\draw[thick,wave](-0.5,\radius)--+(0,\length);

\draw[thick,wave](-1,\radius)--+(0,\length);

\draw[->](2.5,0)--(3,0);

\node[rotate=-45,anchor=west] at(0.2,\radius) {\scriptsize $\bs\sigma^{1}$};
\node[rotate=-45,anchor=west] at(-0.4,\radius) {\scriptsize $\bs\sigma^{0}-\bs h_i+\bs h_j$};
\node[rotate=-45,anchor=west] at(-1,\radius) {\scriptsize $\bs\sigma^{0}-\bs h_i$};
\node[rotate=-45,anchor=west] at(-1.6,\radius) {\scriptsize $\bs\sigma^{0}$};

\end{scope}

\begin{scope}[xshift=5.5cm]

\draw[rounded corners=1cm*\scale, thick] (0,\radius)--(-\radius-\hlength,\radius)--(-\radius-\hlength,-\radius)--
(\radius+\hlength,-\radius)--(\radius+\hlength,\radius)--(0,\radius);


\draw[thick](0,\radius)--+(0,\length);
\draw[thick](0.5,\radius)--+(0,\length);
\draw[thick,wave](-0.5,\radius)--+(0,\length);

\draw[thick,wave](1,\radius)--+(0,\length);

\draw[dashed,->] (1,\length+\radius+0.1) ..controls (0.5,\length+1.5*\radius) and (-0.5,\length+1.5*\radius).. (-1,\length+\radius+0.1);

\draw[->](2.5,0)--(3,0);

\node[rotate=-45,anchor=west] at(0.2,\radius) {\scriptsize $\bs\sigma^{1}$};
\node[rotate=-45,anchor=west] at(-0.4,\radius) {\scriptsize $\bs\sigma^{0}-\bs h_i+\bs h_j$};
\node[rotate=-45,anchor=west] at(-1.6,\radius) {\scriptsize $\bs\sigma^{0}-\bs h_i$};
\node[rotate=-45,anchor=west] at(0.7,\radius) {\scriptsize $\bs\sigma^{0}$};

\end{scope}

\begin{scope}[xshift=11cm]

\draw[rounded corners=1cm*\scale, thick] (0,\radius)--(-\radius-\hlength,\radius)--(-\radius-\hlength,-\radius)--
(\radius+\hlength,-\radius)--(\radius+\hlength,\radius)--(0,\radius);


\draw[thick](0,\radius)--+(0,\length);
\draw[thick](0.5,\radius)--+(0,\length);
\draw[thick,wave](-0.5,\radius)--+(0,\length);

\draw[thick,wave](-1,\radius)--+(0,\length);


\node[rotate=-45,anchor=west] at(0.2,\radius) {\scriptsize $\bs\sigma^{1}-\bs h_k$};
\node[rotate=-45,anchor=west] at(-0.4,\radius) {\scriptsize $\bs\sigma^{0}-\bs h_i+\bs h_j-\bs h_l$};
\node[rotate=-45,anchor=west] at(-1.6,\radius) {\scriptsize $\bs\sigma^{0}-\bs h_i$};
\node[rotate=-45,anchor=west] at(-0.9,\radius) {\scriptsize $\bs\sigma^{0}-\bs h_i-\bs h_l$};

\end{scope}

\end{scope}

\end{tikzpicture}

\end{center}

\caption{Monodromy of degenerate field.}
\label{fig:monodromy}

\end{figure}
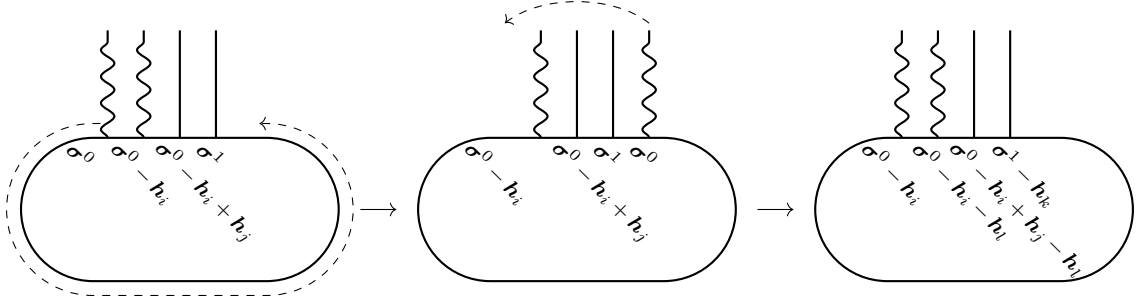

It is a local transformation of conformal blocks, and maps a conformal block to a linear combination of other conformal blocks with different intermediate dimensions.
Since it is local, it can be studied for conformal blocks with one degenerate, one semi-degenerate and two arbitrary fields:
in this case the conformal block is given by a generalized hypergeometric function $_{N}F_{N-1}$,
so the computation of the fusion matrix $F$ is equivalent to re-expansion of hypergeometric function around zero in the vicinity of infinity,
see \cite{Gavrylenko:2018ckn} and references therein.
The analytic continuation between these two region is performed around a semidegenerate field insertion in the counterclockwise direction. These conformal blocks can be obtained directly from geometric engineering in topological string theory, as in \cite{Bonelli:2011fq,Bonelli:2011wx}. We perform the sequence of braiding transformations that correspond to the B-cycle monodromy pictorially, exemplified in the case of two punctures, in Fig.~\ref{fig:monodromy}.

From the figure we can see that after analytic continuation along the B-cycle, the intermediate charges are shifted: in other words, we have an operator-valued monodromy matrix $\hat M_B$, containing shift operators.
The main problem, as in \cite{Iorgov:2014vla,Gavrylenko:2018ckn} will be to turn this matrix into number-valued matrix $M_B$.
Before going through the whole computation let us make the following observation: while in the spherical case all monodromies led to shifts in the $A_{N-1}$ root lattice (generated by $\bs h_i-\bs h_j$),
in the toric case the single B-cycle monodromy also simultaneously shifts all the charges by a single $\bs h_i$.
Therefore the arbitrary shift vector, which appears here and will have to appear in the Fourier transform, has the form $(\bs w^0+\bs\omega_k,\bs w^1+\bs\omega_k,\ldots, \bs w^{n-1}+\bs\omega_k)$,
where $\bs w^l\in Q_{A_{N-1}}$ are the elements of $A_{N-1}$ root lattice.
To get the proper kernel for the Riemann-Hilbert problem it will be necessary to sum over this set: the essential difference from the naive expectation is the presence of the extra shift by the fundamental weight $\bs\omega_k$.

Now we perform the precise computations along the lines of \cite{Gavrylenko:2018ckn}. 
The explicit formula for the fusion kernel is given by
\eq{
{\sf B}_{lj}(\bs\sigma',\nu,\bs\sigma)=e^{-i\pi((N-1)/N+\sigma_l-\sigma'_j)}\prod_{k\neq l}\frac{\sin\pi((\nu+1)/N+\sigma'_j-\sigma_k)}{\sin\pi(\sigma_k-\sigma_l)}\times e^{i\pi N((\bs\sigma+\bs h_l,\bs h_l)-(\bs\sigma_j',\bs h_l))}\,,
}
where the last factor comes from the renormalization of the structure constants \eqref{eq:degenerate_norm} between \cite{Gavrylenko:2018ckn} and the present work. The latter formula can be rewritten in a more compact form:
\eq{\label{eq:DegFusionKer}
{\sf B}_{lj}(\bs\sigma',\nu,\bs\sigma)=e^{\pi i(\nu+1/N)}\prod_{k\neq l}\frac{1-e^{-2\pi i((\nu+1)/N+\sigma_j'-\sigma_k)}}{1-e^{-2\pi i(\sigma_l-\sigma_k)}}\,.
  }
The main advantage of the normalization \eqref{eq:degenerate_norm} is that the new braiding matrix is periodic under $\sigma_i\mapsto\sigma_i+1$ or $\sigma_i'\mapsto\sigma_i'+1$.
  
In matrix notation, the braiding of Figure \ref{fig:fusion} takes the form
\eq{
\mc P_{\bs\sigma'}V_\nu(z)\vec\phi(\gamma\cdot y)\mc P_{\bs\sigma}=
{\sf B}(\bs\sigma',\nu,\bs\sigma)\cdot\mc P_{\bs\sigma'}\vec\phi(y)\Phi(z)\mc P_{\bs\sigma}\,,
}
Another basic operation is the permutation of a degenerate field and a projector:
\eq{
\vec\phi(z)\mc P_{\bs\sigma}=\nabla_{\bs\sigma}\mc P_{\bs\sigma}\vec\phi(z)\,.
}
Here $\nabla_{\bs\sigma}$ is a diagonal matrix with entries given by the shift operators: $(\nabla_{\bs\sigma})_{ii}\mc P_{\bs\sigma}=\mc P_{\bs\sigma+\bs h_i}$. The appearance of such operators makes monodromy matrices operator-valued. The transformation of the conformal block \eqref{eq:toric_block} when we analytically continue in $z$ along the B-cycle is expressed as a sequence of these operations: in order to write it down, it is convenient to introduce the column vectors
\eq{
\vec \Psi_j(z)=\left(\Psi_{1,j},\ldots,\Psi_{N,j}\right)^T\,,
}
constructed from the lines of $\Psi$. In terms of these, we can write the monodromy transformation as
\eq{
\vec\Psi_j(\gamma_B\cdot z)=\hat M_B^T\vec\Psi_j(z)\,,
}
where
\eq{
\hat M_{B}^{T}=
\nabla^{-1}_{\bs\sigma^0} {\sf B}(\bs\sigma^{n-1},\nu_n,\bs\sigma^0)
\nabla^{-1}_{\bs\sigma^{n-1}} {\sf B}(\bs\sigma^{n-2},\nu_{n-1},\bs\sigma^{n-1})
\ldots
\nabla^{-1}_{\bs\sigma^1} {\sf B}(\bs\sigma^0+\bs h_j,\nu_{1},\bs\sigma^1)e^{\pi i(1-N)/N},
}
To compute braiding of two degenerate fields we used the simple identity
\eq{\label{eq:trivialtransf}
{\sf B}(\bs\sigma^0,-1,\bs\sigma^0+\bs h_j-\bs h_l)_{lk}=e^{\pi i(1-N)/N}.
}
To further simplify the form of the monodromy matrix $\hat M_B$ we do some manipulations in order to make all shift operators act only on the conformal blocks, but not on the other matrices. We will denote a shift operator that acts only on the conformal block by $\tilde{\nabla}$.
This can be done with the help of the following identities:
\eq{
\label{eq:shifts}
{\sf B}(\bs\sigma',\nu,\bs\sigma\pm\bs h_m)=-{\sf B}(\bs\sigma',\nu\pm1,\bs\sigma)\,,\\
{\sf B}(\bs\sigma'\pm\bs h_m,\nu,\bs\sigma)=-{\sf B}(\bs\sigma',\nu\mp1,\bs\sigma)
}
and their obvious consequence:
\eq{
\label{eq:nabla_tilde}
\nabla^{-1}_{\bs\sigma}{\sf B}(\bs\sigma',\nu,\bs\sigma)=-\tilde\nabla^{-1}_{\bs\sigma} {\sf B}(\bs\sigma',\nu+1,\bs\sigma).
}
Naively one might think that $\hat M_B^T$ acts differently on different rows of $\Psi$, but due to \eqref{eq:shifts} this dependence disappears.
Simplified form of the monodromy matrix is
\eq{
\hat M_{B}^{T}=
(-1)^ne^{\pi i(1-N)/N}\tilde\nabla^{-1}_{\bs\sigma^0} {\sf B}(\bs\sigma^{n-1},\nu_n-1,\bs\sigma^0)
\tilde\nabla^{-1}_{\bs\sigma^{n-1}} {\sf B}(\bs\sigma^{n-2},\nu_{n-1}-1,\bs\sigma^{n-1})
\ldots\\\ldots
\tilde\nabla^{-1}_{\bs\sigma^1} {\sf B}(\bs\sigma^1,\nu_{2}-1,\bs\sigma^2)
\tilde\nabla^{-1}_{\bs\sigma^1} {\sf B}(\bs\sigma^0,\nu_{1}-1,\bs\sigma^1).
}

\subsection{Fourier transformation}

One can easily verify using ~\eqref{eq:shifts} that
\eq{
\nabla_{\bs\sigma^i}\otimes\nabla^{-1}_{\bs\sigma^i}\otimes\hat M_B=
\tilde\nabla_{\bs\sigma^i}\otimes\tilde\nabla^{-1}_{\bs\sigma^i}\otimes\hat M_B\,,\\
\nabla_{\bs\sigma^0}\otimes\nabla_{\bs\sigma^1}\otimes\cdots\otimes\nabla_{\bs\sigma^{n-1}}\otimes\hat M_B=
\tilde\nabla_{\bs\sigma^0}\otimes\tilde\nabla_{\bs\sigma^1}\otimes\cdots\otimes\tilde\nabla_{\bs\sigma^{n-1}}\otimes\hat M_B\,.
}
this means that the matrix $\hat M_B$ is periodic with respect to shifts by the vectors $(\bs\omega_j,\ldots,\bs\omega_j)+(\bs w^0,\ldots,\bs w^{n-1})$, where $\bs w^i\in Q_{A_{N-1}}$.

We can thus construct a Fourier transformation of the fundamental solution in order to (almost) diagonalize all shift operators simultaneously:
\eq{
\Psi^D_k\equiv\sum_{\bs w^i\in Q_{A_{N-1}}} e^{2\pi i\sum_{i=0}^{n-1}(\bs\eta^i,\bs w^i+\bs\omega_k)} \Psi\left(\{\bs\sigma^i+\bs w^i+\bs\omega_k\}\right)\,.
}
The shift operators act on this expression as follows:
\eq{
  \nabla^{-1}_{\bs\sigma^0}\otimes \nabla^{-1}_{\bs\sigma^1}\otimes\cdots\otimes \nabla^{-1}_{\bs\sigma^{n-1}} \Psi_k^D=
  e^{2\pi i\tilde{\bs\eta}^0}\otimes e^{2\pi i\tilde{\bs\eta}^1}\otimes\cdots\otimes e^{2\pi i\tilde{\bs\eta}^{n-1}} \Psi_{k-1}^D
}
This means that one can replace
\eq{
\nabla^{-1}_{\bs\sigma^0}\otimes \nabla^{-1}_{\bs\sigma^1}\otimes\cdots\otimes \nabla^{-1}_{\bs\sigma^{n-1}} \to e^{2\pi i\tilde{\bs\eta}^0}\otimes e^{2\pi i \tilde{\bs\eta}^1}\otimes\cdots\otimes e^{2\pi i\tilde{\bs\eta}^{n-1}}T^{-1}\,,
}
where the operator $T$ shifts the index $k\in \mathbb{Z}/N\mathbb{Z}$:
\begin{equation}
T:\Psi_k^D\mapsto \Psi_{k-1}^D.
\end{equation}
Thanks to this, the B-cycle monodromy matrix of $\Psi^D$ is given by
\eq{
\hat M_{B}^{T}=
(-1)^ne^{\pi i(1-N)/N}e^{\tilde{2\pi i\bs\eta}^0} {\sf B}(\bs\sigma^{n-1},\nu_n-1,\bs\sigma^0)
e^{2\pi i\tilde{\bs\eta}^{n-1}} {\sf B}(\bs\sigma^{n-2},\nu_{n-1}-1,\bs\sigma^{n-1})
\ldots\\\ldots
e^{2\pi i\tilde{\bs\eta}^2} {\sf B}(\bs\sigma^1,\nu_{2}-1,\bs\sigma^2)
e^{2\pi i\tilde{\bs\eta}^1} {\sf B}(\bs\sigma^0,\nu_{1}-1,\bs\sigma^1).
}

The A-cycle monodromy can be computed in the obvious way, but the problem is that it is different in the sectors with different shifts $\bs\omega_k$:
\eq{
M_{A,k}=e^{2\pi i(\tilde{\bs\sigma}^0-\bs\omega_k)}=e^{2\pi i k/N}e^{2\pi i\tilde{\bs\sigma}^0}\,.
}
To fix this issue it is necessary to introduce an extra $U(1)$ boson $\varphi(z)$ with the OPE
\eq{
\varphi(z)\varphi(w)\sim -\frac1N\log(z-w)
}
Using this boson we turn W-degenerate fields into N--component fermions:
\eq{
\psi_i(z)=\phi_i(z)\otimes e^{i\varphi(z)}\,,\\
\bar\psi_i(z)=\bar\phi_i(z)\otimes e^{-i\varphi(z)}\,.
}
After analogous, but quite simpler considerations w.r.t. the ones reported above, we arrive at the result that, for the $U(1)$ factor, the B--cycle monodromy is just the charge-shifting operator for $U(1)$ charge, and the A-cycle monodromy is just some number, different in the different sectors:
\eq{
\hat M_B^{U(1)}=e^{2\pi i(\rho+\frac{N-1}{2N})}\left(T^{U(1)}\right)^{-1}\,,\\
M_{A,k}^{U(1)}=e^{-2\pi i k/N}e^{2\pi i \sigma^{U(1)}}\,,
}
where the $U(1)$ shift operator is defined as
\begin{equation}
T^{U(1)} f(\sigma)=f\left(\sigma+\frac1N\right)
\end{equation}

We are finally able to construct the following object, which is invariant under the action of $T\cdot T^{U(1)}$:
\eq{
\Psi^{U(N)}(z,z_0)=\sum_{k=0}^{N-1} \Psi_k^D(z,z_0)\Psi^{U(1)}_k(z,z_0),
}
that has number-valued monodromies:
\eq{\label{eq:VerlindeMonodromyB}
M_{B}^{T}=
(-1)^ne^{{2\pi i\bs\eta}^0} {\sf B}(\bs\sigma^{n-1},\nu_n-1,\bs\sigma^0)
e^{2\pi i{\bs\eta}^{n-1}} {\sf B}(\bs\sigma^{n-2},\nu_{n-1}-1,\bs\sigma^{n-1})
\ldots\\\ldots
e^{2\pi i{\bs\eta}^2} {\sf B}(\bs\sigma^1,\nu_{2}-1,\bs\sigma^2)
e^{2\pi i{\bs\eta}^1} {\sf B}(\bs\sigma^0,\nu_{1}-1,\bs\sigma^1),\\
M_{A}^{U(N)}=e^{2\pi i\bs\sigma^0}\,,
}
giving a solution to the Riemann-Hilbert problem.

Finally, let us note that from this we can read the explicit form of the fermion braiding matrix $B_k$ used in the previous section
:
\begin{equation}\label{eq:BraidSemiDeg}
B_k= - {\sf B}^T(\bs\sigma^{k-2},\nu_{k-1}-1,\bs\sigma^{k-1}) e^{2\pi i{\bs\eta}^{k-1}}.
\end{equation}











\section{Relation to Krichever's connection}\label{sec:KricheverCFT}

We wish now to connect the solution we found in the previous sections to the solution of the linear system defined by the Lax matrix
\begin{align}\label{eq:KricheverLax}
L_{ii}(z|\tau)=p_i+\sum_kL_{ii}^{(k)}\left[\zeta(z-z_k)-\zeta(z-Q_i)-\zeta(Q_i-z_m) \right], && \sum_mL_m^{ii}=-1,
\end{align}
\begin{align}
L_{ij}(z|\tau)=\sum_k L_{ij}^{(k)}\left[\zeta(z-z_k)-\zeta(z-Q_j)-\zeta(Q_i-z_k)+\zeta(Q_i-Q_j) \right], && i\ne j,
\end{align}
obtained following Krichever's construction \cite{Krichever:2001zg,Krichever:2001cx}, which is a different approach to the construction of Lax matrices on elliptic curves, that also extends to algebraic curves of higher genus.

Recall that Riemann-Roch theorem forces, in the $g>0$ case, the introduction of twist factors that we discussed in Section \ref{sec:LinSys}. More specifically, a Lax matrix is a meromorphic matrix-valued differential with poles specified by a divisor on the Riemann surface. The space of $r\times r$ matrix functions with degree $d$ divisor of poles has dimension $r^2(d-g+1)$. Besides the Lax pair matrices $L,M$, the Lax equation involves also their commutator: if $n,m$ are the degrees of the divisors of $L,M$ respectively, the degree of their commutator is $n+m$. We thus have $r^2(n+m-g+1)$ equations, but only $r^2(n+m-2g+1)$ unknown functions modulo gauge equivalence. Unless $g=0$, this results in an overdetermined system of equations. One way of dealing with this is tensoring with some other bundle, which is technically what we do when we introduce twists: our Lax matrix was not a meromorphic differential but rather a section of some other bundle, so we cannot straightforwardly apply Riemann-Roch theorem as above.

There exists another way to handle this problem, which is to consider the linear system as defining a vector bundle of degree $rg$, instead of a degree zero bundle as in the construction with twists. Then the determinant bundle will vanish at $rg$ points, and one can show that the Lax matrix $L(z)$ for such a linear system will have additional simple poles at extra points, the so-called Tyurin points. These simple poles have residue one, so that from the point of view of the linear system they are apparent singularities around which the solution of the linear system
\begin{equation}
\partial_zY^{Kr}(z|\tau)=L^{Kr}(z|\tau)Y^{Kr}(z|\tau)
\end{equation}
will have no monodromies. The Riemann-Hilbert problem for $Y^{Kr}$ is modified as following: instead of having \eqref{eq:Monodromies}, we have
\begin{equation}\label{eq:RHPKrich}
\begin{cases}
Y(\gamma_k\cdot z|\tau)=Y(z|\tau)M_k, & k=1,\dots,n \\ 
Y(z+1|\tau)=Y(z|\tau)M_A, \\
Y(z+\tau|\tau)=Y(z|\tau)M_B,\\
\det Y(Q_i|\tau)=0, & i=1,\dots,r,
\end{cases}
\end{equation}
with Lax matrix given by \eqref{eq:KricheverLax}.

To make contact with our fermionic construction, first recall that in Section \ref{sec:LinSys} it was mentioned that starting from the original Lax matrix \eqref{eq:LaxTorus} it is possible to go to a description involving a different one by means of a singular gauge transformation, so that to find the CFT description of this approach we should find a $g(z|\tau)$ such that
\begin{equation}
L^{Kr}=gLg^{-1}+\partial_z g g^{-1}.
\end{equation}

For the one-punctured torus with a single pole at zero, the Lax matrices are
\begin{gather}
L^{Kr}_{ij}=m\frac{\theta_1(z+Q_i-Q_j)\theta_1(z-Q_i)\theta_1(Q_j)}{\theta_1(z)\theta_1(z-Q_j)\theta_1(Q_i-Q_j)\theta_1(Q_i)}, \\ L_{ii}^{Kr}=p_i+E_1(z-Q_i)-E_1(z)+E_1(Q_i)
\end{gather}
\begin{align}
L^{CM}_{ij}=mx(Q_i-Q_j,z), && L^{CM}_{ii}=p_i,
\end{align}

so that the gauge transformation is relatively easy to find, and is given by
\begin{equation}
g(z)=\diag\left[\frac{\theta_1'(0)\theta_1(z+Q_i)}{\theta_1(z)\theta_1(Q_i)} \right].
\end{equation}
To generalize this to the case of many punctures, it is convenient instead to consider the Riemann-Hilbert Problem \eqref{eq:RHPKrich}. Such solution, that we will from now on denote by $Y^{Kr}$, can be constructed from $Y(z)$ 
in the following way:
\eq{\label{eq:GaugeTransform}
Y^{Kr}(z)=\diag \frac{\theta_1'(0)\theta_1(z-z_1+Q_i)}{\theta_1(z-z_1)\theta_1(Q_i-z_1)}\times Y(z)\equiv g(z)Y(z)
}
We see that $\det Y^{Kr}(z_1-Q_i)=0$ in all points $Q_i$, and also all its singular exponents in the point $z_1$ are shifted.
One way to obtain this solution is the following: consider first the kernel \eqref{eq:RHKernel}:
\eq{
K(z,z_0)=Y(z_0)^{-1}\frac{\theta_1'(0)\theta_1(z-z_0+\bs Q)}{\theta_1(z-z_0)\theta_1(\bs Q)}Y(z),
}
and then send $z_0\to z_1$:
\eq{
Y^{Kr}(z)=\lim_{z_0\to z_1} Y(z_0)K(z,z_0).
\label{eq:Krichever_limit}
}
This formula has clear CFT interpretation: near $z_1$ the behavior of the solution is
\eq{
Y(z_0)=G_1(z_0-z_1)(z_0-z_1)^{\bs\theta_1}C_1,
}
where $G_1(z)$ is holomorphic and invertible around $z=0$. Therefore
\eq{
K(z,z_0)=C_1^{-1}(z_0-z_1)^{-\bs\theta_1}G_1(z_0-z_1)^{-1}Y^{Kr}(z).
}
Because of the limit $z_0\rightarrow z_1$, in the CFT we have to consider the OPE of the fermion $\bar\psi_\alpha(z_0)$ with the primary field $V_{\bs\theta_1}(z_1)$:
\eq{
\bar\psi_\alpha(z_0)V_{\bs\theta_1}(z_1)=\sum_\beta\left(C_1^{-1}\right)_{\alpha\beta}(z_0-z_1)^{-\theta_{1,\beta}}\delta_\beta V_{\bs\theta_1}(z_1)+\ldots,
\label{eq:OPE_with_fermion}
}
where $\delta_\beta V_{\bs\theta_1}$ is a field with shifted W-charge $\bs\theta_1\mapsto\bs\theta_1-\bs h_\beta$~\footnote{Notice
that in the general $W_N$ case fields $\delta_\beta V$ are rather problematic. The only well-understood fields are the ones with $\bs\theta_1=\nu_1\bs\omega_1$, but fields with charge $\bs\theta_1-\bs h_\beta$ generally do not lie in this class.}.

Now comparing \eqref{eq:OPE_with_fermion} with \eqref{eq:Krichever_limit} we can identify
\eq{
Y^{Kr}(z)_{\alpha\beta}=\sum_\gamma G_1(0)_{\alpha\gamma}\frac{\langle\psi_\alpha(z)
\delta_\gamma V_{\bs\theta_1}(z_1)V_{\bs\theta_2}(z_2)\ldots V_{\bs\theta_n}(z_n)\rangle}{\langle V_{\bs\theta_1}(z_1)\ldots V_{\bs\theta_n}(z_n)\rangle}.
}
We see that up to normalization (which is not actually fixed) Krichever's solution has the nice CFT interpretation of the expectation value of the single fermion in the presence of all the vertex operators. The expression of the two-fermionic correlator in terms of Krichever's solution can be obtained by applying the gauge transformation \eqref{eq:GaugeTransform}:
\eq{
K(z,z_0)=Y(z_0)^{-1}\frac{\theta_1'(0)\theta_1(z-z_0+\bs Q)}{\theta_1(z-z_0)\theta_1(\bs Q)}Y(z)=
\\
=Y^{Kr}(z_0)^{-1}\frac{\theta_1(z_0-a_1+\bs Q)}{\theta_1(z_0-a_1)}\frac{\theta_1'(0)\theta_1(z-z_0+\bs Q)}{\theta_1(z-z_0)\theta_1(\bs Q)}\frac{\theta_1(z-a_1)}{\theta_1(z-a_1+\bs Q)}Y^{Kr}(z).
}
We thus see that Krichever's solution becomes less natural than $Y(z)$ if we wish to express the two-fermionic correlator, because it contains a more involved diagonal matrix between $Y^{Kr}(z_0)^{-1}$ and $Y^{Kr}(z)$. On the other hand, contrary to what happens in the twisted formulation, the solution itself can be obtained from the CFT, not only the kernel.

\section{Solution of the elliptic Schlesinger system} \label{sec:EllGaudin}

As a further application of our results 
we will now show how, starting from equations \eqref{eq:TauFunctionCFT}, \eqref{eq:TauFunctionGaugeU}, one can obtain a formula for the solution of the Calogero-like variables $Q_i$ of the elliptic Schlesinger system. This formula generalizes the algebro-geometric solution of the elliptic Calogero-Moser model found in \cite{1999JMP....40.6339G} 
to the nonautonomous case with many punctures, and suggests a double role of  the dual partition function from the point of view of integrable systems: on the one hand, being proportional to the tau function, its vanishing locus includes the Malgrange divisor, where the Riemann-Hilbert problem is no longer solvable \cite{Malgrange1982,2010CMaPh.294..539B,2016arXiv160104790B}. On the other hand, we have an extra vanishing locus, which generalizes the Riemann theta divisor of the Krichever/Seiberg-Witten curve, whose points are the solution to the equations of motion of the isomonodromic system. Note that this is essentially a consequence of our choice of twists, or analogously of the choice of Calogero-like dynamical variables. As a byproduct, we will also obtain a direct link between the isomonodromic tau function and the $SU(n)$, rather than $U(n)$, gauge theory.

However, note that in the case of more than one puncture, the Calogero-like variables $Q_i$ do not specify the whole system: there are additional spin variables satisfying the Kirillov-Kostant Poisson bracket  for $\mathfrak{sl}(N)$ \cite{Takasaki:2001fr} that will not enter in the following discussion: while it may be that there is some further connection between $Z^D$ and these remaining dynamical variables, this does not seem evident at the moment.

In order to obtain the aforementioned result, we first split $Z^D$ in various components having different types of ${\mathfrak gl}_N$ shifts:
\begin{equation}
Z^D=\tr_{\mathcal{H}}(-)^Fe^{2\pi i\bs\eta\cdot\bs J_0}q^{L_0}V=\sum_{\textbf{n}\in\mathbb{Z}^N}\tr_{\mathcal{H}_{\textbf{n}}}(-)^Fe^{2\pi i\bs\eta\cdot\bs J_0}q^{L_0}V,
\end{equation}
where we denoted by $V$ the whole string of vertex operators. To perform the splitting, it is convenient to decompose $\bs\eta$ as
\begin{align}
\bs\eta=\eta_1\bs\omega_1+\dots+\eta_{N-1}\bs\omega_{N-1}+N\rho\textbf{e}\equiv\bs{\tilde{\eta}}+N\rho\textbf{e},
\end{align}
where
\begin{equation}
\textbf{e}\equiv \frac{1}{N}\left(1,\dots,1 \right),
\end{equation}
and $\bs\omega_k$ are the fundamental weights of $\mathfrak{sl}_N$, normalized as
\begin{equation}
\bs\omega_k\cdot\bs\omega_k=k\frac{N-k}{N}.
\end{equation}
We also decompose \textbf{n} as
\begin{equation}
\textbf{n}=(n_1,\dots,n_N)\equiv \tilde{\textbf{n}}+N\left(k+\frac{j}{N} \right) \textbf{e} ,
\end{equation}
where we separated the traceless part from the $U(1)$ factor 
\begin{equation}
J_0^{U(1)}=\textbf{n}\cdot\textbf{e}=\frac{n}{N}\equiv k+\frac{j}{N},
\end{equation}
with $j=0,\dots,N-1$. The space $\mathcal{H}_{\textbf{n}}$ analogously decomposes into a $W_N$ highest weight module plus a Fock space, with $U(1)$ charge given by
\begin{equation}
\mathcal{H}_{\textbf{n}}=\mathcal{W}_{\textbf{a}+\bs{\tilde{n}}}\oplus\mathfrak{F}_{\sigma+1/2+k+j/N},
\end{equation}
where we shifted the $U(1)$ charge $\sigma$ by $1/2$ to get consistent signs in the monodromy, as in the $2\times 2$ case. Then,
\begin{equation}
\begin{split}
Z^D & =\sum_{j=0}^{N-1}\sum_{k,\tilde{\textbf{n}}}\tr_{\mathcal{W}_{\textbf{a}+\tilde{\textbf{n}}}}\left(e^{2\pi i\bs\eta\cdot\tilde{\textbf{n}}}q^{L_0}V\right)\tr_{\mathfrak{F}_{\sigma+k+j/N+1/2}}\left(e^{2\pi iN(\rho+1/2)(k+j/N)}q^{L_0} \right), \\
\end{split}
\end{equation}
where we encoded the fermion number operator into a shift of $\rho$ by $1/2$. However, we must note that $j$ is not independent of the $W_N$ charge shift. In fact, if we parametrize
\begin{equation}
\textbf{n}=(n_1+k,n_2+k,\dots,n_{N-1}+k,k),
\end{equation}
the $U(1)$ charge is indeed
\begin{equation}
J_0^{U(1)}=k+\frac{n_1+\dots+n_{N-1}}{N}\equiv k+\frac{j}{N},
\end{equation}
but we also have
\begin{equation}
\frac{n_1+\dots n_{N-1}}{N}=\textbf{n}\cdot\bs\omega_{N-1},
\end{equation}
so that $j/N$ is the shift in the $W_N$ weight along the $\bs\omega_{N-1}$ direction. Then,
\begin{equation}\label{ZDDecomp}
\begin{split}
Z^D& =\frac{1}{\eta(\tau)} \sum_{j=0}^{N-1}Z^D_j\sum_ke^{2\pi iN(\rho+1/2)(k+j/N)}e^{N\pi i\tau(\sigma+k+j/N+1/2)^2}\\
& =\frac{q^{\sigma^2}q^{N/8}q^{N(\sigma\tau+1/2)/2}}{\eta(\tau)}\sum_{j=0}^{N-1}\theta_{N\tau}\left[ \begin{array}{c}
j/N \\ 0
\end{array} \right](N(\rho+1/2+(\sigma+1/2)\tau))Z^D_j,
\end{split}
\end{equation}
where we defined
\begin{equation}
Z_j^D\equiv \sum_{\tilde{\bs n}\in Q_{A_{N-1}},\,\tilde{\bs n}\cdot\bs\omega_{N-1}=j/N}\tr_{\mathcal{W}_{\bs a+\tilde{\bs n}}}e^{2\pi i\bs\eta\cdot\tilde{\bs n}}q^{L_0}V
\end{equation}
We should now compare with
\begin{equation}
Z^D=\Tau\prod_i\frac{\theta_1(Q_i-\sigma\tau-\rho)}{\eta(\tau)}.
\end{equation}
First of all, from this expression we see that $\sigma\tau+\rho=Q_i$ are zeros of $Z_D$. In other words, the solutions of the nonautonomous system are given by
\begin{align}
Z^D|_{\sigma\tau+\rho=Q_i}=0, && i=1,\dots,N.
\end{align}
This is a generalization to the nonautonomous case of the condition $\theta(Q)=0$, expressing the solution of the autonomous integrable system as the vanishing theta divisor of the Seiberg-Witten curve, which is the autonomous limit of our description. Further, the decomposition \eqref{ZDDecomp} is a deformation of the one expressing the Riemann theta function associated to the Seiberg-Witten curve as a sum over $N-1$ Jacobi theta functions with characteristics shifted by $j$ \cite{1999JMP....40.6339G}.

We can further write the isomonodromic tau function in a way that is manifestly independent from the $U(1)$ charges. By writing all the theta functions in their $q$-series representation, we have
\begin{equation}
\begin{split}
Z^D & =\frac{q^{\sigma^2}}{\eta(\tau)^N}(i)^N \Tau \sum_{n_1,\dots,n_N}(-)^{n_1+\dots n_N}e^{2\pi i\tau\left[\left(n_1+1/2\right)^2+\dots+\left(n_N+1/2 \right)^2 \right]/2} \\
& \times e^{2\pi i\left[\left(n_1+1/2 \right)(-Q_1+\sigma\tau+\rho)+\dots+\left(n_N+1/2 \right)(-Q_N+\sigma\tau+\rho) \right]} \\
& =\frac{q^{\sigma^2}}{\eta(\tau)^N}(i)^N\Tau\sum_{\textbf{n}\in\mathbb{Z}^N}e^{2\pi i \textbf{n}\cdot\left(-\textbf{Q}+(\sigma\tau+\rho+1/2)\textbf{e} \right)}e^{2\pi i (\textbf{n}+\textbf{e}/2)^2\tau/2}.
\end{split}
\end{equation}
We decompose, similarly as before,
\begin{equation}
\textbf{n}=\tilde{\textbf{n}}+N(n+j/N)\textbf{e},
\end{equation}
and find
\begin{equation}
\begin{split}
Z^D & =i^N\Tau\frac{q^{\sigma^2}}{\eta(\tau)^N}(i)^Ne^{i\pi N(\sigma\tau+\rho)} \sum_{j=0}^{N-1}\sum_{\tilde{\textbf{n}},k}\left(e^{2\pi i\tilde{\textbf{n}}\cdot\textbf{Q}}e^{i\pi\tilde{\textbf{n}}^2\tau} \right)\left(e^{i\pi N(k+j/N)}e^{i\pi N\tau(k+j/N+1/2)^2}e^{2\pi i N(\sigma\tau+\rho)(k+j/N)} \right) \\
& =\Tau\frac{q^{\sigma^2}}{\eta(\tau)^N} e^{i\pi N(\sigma\tau+\rho+1/2)}\sum_{j=0}^{N-1}\Theta_j(\textbf{Q})\sum_ke^{i\pi N\tau(k+j/N)^2}e^{2\pi iN\left((\sigma+1/2)\tau+\rho+1/2\right)} \\
& =\Tau\frac{q^{\sigma^2}}{\eta(\tau)^N}e^{i\pi N\left(\sigma\tau+\rho+1/2\right)}\sum_{j=0}^{N-1}\Theta_j(\textbf{Q})\theta_{N\tau}\left[ \begin{array}{c} j/N \\ 0
\end{array}\right]\left(N((\sigma+1/2)\tau+\rho+1/2)\right).
\end{split}
\end{equation}
Comparing the two expressions, we see that
\begin{equation}\label{miononno}
Z^D_j=\frac{e^{i\pi N\rho}}{\eta(\tau)^{N-1}}\Theta_j(\textbf{Q})\Tau,
\end{equation}
where
\begin{equation}\label{eq:RiemannTheta}
\Theta_j(\textbf{Q})=\sum_{ \textbf{n}\in Q_{A_{N-1}}:\text{ }\textbf{n}\cdot\bs\omega_{N-1}=j/N}e^{2\pi i\textbf{n}\cdot \textbf{Q}}e^{i\pi\textbf{n}^2\tau}.
\end{equation}

\section{Conclusions and outlook}\label{sec:Concl}

In this paper we showed how the isomonodromic tau function for a linear system on the torus with $n$ regular singularities can be expressed as a Fourier transform of conformal blocks in a free fermionic CFT, where the Fourier transform is obtained by summing over all the fermion charges under the Cartan of a twisted $\widehat{\mathfrak{gl}}(N)_1 $ algebra. Through the AGT correspondence, these are related in the usual way to dual partition functions of a circular quiver gauge theory, so that this results extends the Painlev\'e/gauge theory correspondence \cite{Gamayun:2013auu,Bonelli:2016qwg} to the case of circular quiver theories in class $\mathcal{S}$ with gauge groups $SU(N)$ and their S-duals, obtained by wrapping a stack of N M5-branes on a punctured torus with an arbitrary number of punctures. Let us remark that as a consequence of this identification we get a relation between the solution of the multi particle integrable deautonomized system and the gauge theory dual partition functions
$Z^D_j$, which can be regarded alternatively as equations for the gauge theory partition functions given the solution of the integrable system \eqref{miononno}.

An interesting direction for further studies is the relation to surface operators in gauge theory: in class $\mathcal{S}$ theories there are two ways of constructing surface operators\cite{Gaiotto:2009fs}: from intersecting another set of M5 branes with the original ones wrapping the Riemann Surface that define the theory (codimension 2 defects) or from M2 branes with endpoints on the original M5s (codimension four defects). This latter type of surface operator is localized at one point on the Riemann surface: in the context of the AGT correspondence the partition function in the presence of such a surface operator is realized by the insertion of a Virasoro degenerate or $W_N$ completely degenerate field in the conformal block that yields the usual instanton partition function \cite{Alday:2009fs,Drukker:2009id}. As we show in the appendix, our fermions are constructed from such degenerate fields just by adding a $U(1)$ boson: as such, the kernel \eqref{eq:RHKernel} is naturally related to such objects. The other possible surface operator instead has dimension four in the six-dimensional theory, and wraps the whole Riemann surface: the relation between these two types of surface operators, also from the CFT viewpoint, has been discussed in \cite{Frenkel:2015rda}. In the 2d CFT, this amounts to changing the theory itself, and the partition function in the presence of such a surface operator is given by a conformal block of a $\widehat{\mathfrak{sl}}(N)_k$ algebra with an insertion of a certain twist operator $\mathcal{K}$ -- see \cite{Alday:2010vg}, with level $k$ related to the equivariant parameters by requiring that the original Virasoro algebra of the Liouville theory is recovered upon quantum Drinfeld-Sokolov reduction, i.e.
\begin{equation}
k=-N-\frac{\epsilon_2}{\epsilon_1}.
\end{equation}
Because of this, the partition function in the presence of a codimension four surface operator is a solution of KZB equations \cite{Bernard:1987df,Bernard:1988yv}, which are known to be a quantization of isomonodromy deformation equations \cite{Reshetikhin1992,Harnad:1994fk}. In fact it is expected
\cite{NekrasovSeminar2019} that the classical $k\rightarrow\infty$ limit of the partition function with codimension four surface defect reproduces the formula identifying the tau function with the dual gauge theory partition function.
On the one hand, it would be interesting to investigate how the extra factors present in our formulas arise when doing such a procedure on a circular quiver theory, or even more simply in the $\mathcal{N}=2^*$ theory. On the other hand, it should be noted that we already have a (\textit{twisted}) Kac-Moody algebra in our construction, but with fixed level one. The relation between the appearance of a twisted KM algebra at level one and that of the classical limit of an untwisted KM algebra with the additional insertion of a twist operator $\mathcal{K}$ certainly needs further elucidation. Moreover, it would be interesting to lift our analysis to 5d SUSY gauge theories and group Hitchin systems \cite{Elliott:2018yqm}, but on elliptic curves, in which case discrete Painlev\'e equations should play a central role.

\quad

\quad

{\bf Acknowledgments}: We would like to thank M. Bertola, J. Harnad, D. Korotkin for interesting discussions.
The work of G.B. and F.D.M. is partially supported by INFN - ST\&FI. The work of P.G. was carried out within the HSE University Basic Research Program and funded by the Russian Academic Excellence Project '5-100'. The results of Section \ref{sec:VerlindeSol} were obtained under the support of  Russian science foundation within the  grant 19-11-00275. The work of  A.T. is partially supported by INFN - GAST. The work of A.T. is supported by PRIN project "Geometria delle variet\`a algebriche". The work of G.B. is supported by the PRIN project "Non-perturbative Aspects Of Gauge Theories And Strings".

\begin{appendix}
\section{Elliptic and theta functions}\label{sec:Theta}
We consider tori normalized so that their periods are $(1,\tau)$. In our discussion appear two different theta functions: the Jacobi theta function with characteristics
\begin{equation}
\theta_\tau\left[ \begin{array}{cc} a \\ b
\end{array} \right](z)=\sum_{n\in\mathbb{Z}}e^{i\pi(n+a)^2\tau}e^{2\pi i(z+b)(n+a)}
\end{equation}
and its specification
\eq{
\begin{split}
&\theta_1(z|\tau) \equiv \theta_\tau\left[ \begin{array}{cc} \frac{1}{2} \\ \frac{1}{2}
\end{array} \right](z)= -i\sum_{n\in\mathbb{Z}}(-1)^n e^{i\pi(n+\frac12)^2\tau} e^{2\pi iz(n+\frac12)}.
\end{split}
}
In the following we will also denote 
$
q=e^{2\pi i\tau}$.
A $z$-derivative is indicated by a prime, and when the theta function or its derivatives are evaluated at $z=0$, we simply omit the $z$-dependence: $\theta_1(0|\tau)\equiv\theta_1(\tau)$, and so on. The quasi-periodicity properties of the theta functions are
\begin{align}
\theta_\tau\left[ \begin{array}{cc} a \\ b
\end{array} \right](z+1)=e^{2\pi ia}\theta_\tau\left[ \begin{array}{cc} a \\ b
\end{array} \right](z), && \theta_\tau\left[ \begin{array}{cc} a \\ b
\end{array} \right](z+\tau)=e^{-i\pi\tau-2\pi i z-2\pi i b}\theta_\tau\left[ \begin{array}{cc} a \\ b
\end{array} \right](z),
\end{align}
so that
\begin{align}\label{eq:ThetaPeriod}
\theta_1(z+1|\tau)=-\theta_1(z|\tau), && \theta_1(z+\tau|\tau)=-q^{-1/2}e^{-2\pi i z}\theta_1(z|\tau).
\end{align}

We also use Weierstrass elliptic functions $\wp$ and $\zeta$. $\wp$ is an elliptic function with a single double pole at $z=0$. Its expression in terms of theta functions is
\begin{equation}
\wp(z|\tau)=-\partial_z^2\log\theta_1(z|\tau)-2\eta_1(\tau)=-\zeta'(z|\tau),
\end{equation}
where
\begin{equation}
\eta_1(\tau)=-\frac{1}{6}\frac{\theta_1'''(\tau)}{\theta_1'(\tau)}.
\end{equation}
Weierstrass' $\zeta$ function is minus the primitive of $\wp$. It has only one simple pole at $z=0$, with an affine quasi-periodicity along the A- and B-cycle:
\begin{equation}
\zeta(z|\tau)= 2\eta_1(\tau)z+\partial_z\log\theta_1(z|\tau),
\end{equation}
\begin{align}
\zeta(z+1|\tau)=\zeta(z|\tau)+ 2\eta_1(\tau), && \zeta(z+\tau|\tau)=\zeta(z|\tau)+ 2\tau\eta_1(\tau)-2\pi i.
\end{align}
It turns out to be convenient to normalize the Weierstrass elliptic functions in a different way, in order to have vanishing A-cycle integral. The functions thus obtained are called Eisenstein functions:
\begin{gather}\label{eq:Eisenstein}
E_1(z|\tau)=\partial_z\log\theta(z|\tau)=\zeta(z|\tau)-2\eta_1(\tau)z, \\
E_2(z|\tau)=-\partial_z E_1(z|\tau)=\wp(z|\tau)+2\eta_1(\tau).
\end{gather}
Finally, we use Dedekind's $\eta$ function, defined as
\begin{equation}
\eta(\tau)=q^{1/24}\prod_{n=1}^\infty(1-q^n).
\end{equation}
It is related to the function $\theta_1$ by
\begin{equation}
\eta(\tau)=\left(\frac{\theta_1'(\tau)}{2\pi} \right)^{1/3}.
\end{equation}

Because of the quasi-periodicity properties of the theta functions \eqref{eq:ThetaPeriod}, the Lam\'e function
\begin{equation}
x(u,z)=\frac{\theta_1(z-u|\tau)\theta_1'(\tau)}{\theta_1(z|\tau)\theta_1(u|\tau)}
\end{equation}
has the following transformations:
\begin{align}
x(u,z+1)=x(u,z), && x(u,z+\tau)=e^{2\pi i u}x(u,z).
\end{align}
$x(u,z)$ has the following important property:
\begin{equation}\label{eq:LameProp0}
2\pi i\partial_\tau x(u,z)+\partial_z\partial_u x(u,z)=0.
\end{equation}

\section{$W_N$ algebra and degenerate fields}\label{sec:WNApp}

$W_N$ algebras, first introduced by Zamolodchikov \cite{Zamolodchikov:1985wn}, are infinite-dimensional algebras with generators up to spin $N$ \cite{Bouwknegt:1988sv,Kausch:1990bn,Blumenhagen:1990jv}. They are a higher-spin generalization of the Virasoro algebra, which is the particular case $W_2$, generated by the energy-momentum tensor $T(z)$ of spin 2. In this Appendix we provide the necessary definitions for the $c=N-1$ $W_N$ algebra generators, conformal blocks, and degenerate fields, as well as showing the connection between degenerate fields and free fermions. In the following we will use fundamental weights plus zero vector given by
\eq{
\bs\omega_0=(0,0,0,\ldots,0)\,,\\
\bs\omega_1=(\frac{N-1}{N},\frac{-1}{N},\frac{-1}{N},\ldots,\frac{-1}{N})\,,\\
\bs\omega_2=(\frac{N-2}{N},\frac{N-2}{N},\frac{-2}{N},\ldots,\frac{-2}{N})\,,\\
\ldots\\
\bs\omega_{N-1}=(\frac{1}{N},\frac{1}{N},\frac{1}{N},\ldots,\frac{1-N}{N})\,,\\
}
to be distinguished from the weights of the first fundamental representation of $\mathfrak{sl}_N$:
\eq{
\bs h_1=(\frac{N-1}{N},\frac{-1}{N},\frac{-1}{N},\ldots,\frac{-1}{N})\,,\\
\bs h_2=(\frac{-1}{N},\frac{N-1}{N},\frac{-1}{N},\ldots,\frac{-1}{N})\,,\\
\ldots\\
\bs h_N=(\frac{-1}{N},\frac{-1}{N},\frac{-1}{N},\ldots,\frac{N-1}{N})\,.\\
}

A $W_N$ algebra can be embedded in a $\widehat{\mathfrak{sl}}_N$ algebra, and in fact a $W_N$ CFT can be represented as a constrained WZNW model through the so-called quantum Drinfeld-Sokolov reduction \cite{Bershadsky:1989mf,Forgacs:1989ac}. In particular, for $c=N-1$ there is a realization of the $W_N$ algebra in terms of free bosons $\varphi_k$, subject to the relation
\begin{equation}
\sum_{k=1}^N\varphi_k=0.
\end{equation}
The $W_N$ algebra generators are defined in terms of the $U(1)$ currents generated by these free bosons:
\begin{align}
J_k=i\partial\varphi_k, && W^{(j)}=\sum_{1\le i_1\le N}:J_{i_1}\dots J_{i_j}:,
\end{align}
where $j=2,\dots N$. In particular, note that $W^{(2)}$ is the Sugawara energy-momentum tensor associated to the current algebra.

Analogously to the case of Virasoro, where we can find a basis of the Verma module $\mathcal{V}_{\boldsymbol\theta}$ labeled by partitions, in the $W_N$ case we can find a basis labeled by $N-1$-tuples of partitions $\lambda^{(j)}=(\lambda_1^{(j)},\dots,\lambda_k^{(j)}) $, given by
\begin{equation}
|\boldsymbol\lambda,\boldsymbol\theta\rangle \equiv W^{(N)}_{-\lambda^{(N)}}\dots W_{-\lambda^{(2)}}^{(2)}|\boldsymbol\theta\rangle\equiv W_{\boldsymbol\lambda}|\boldsymbol\theta\rangle,
\end{equation}
where $W_{-\lambda}^{(j)}$ represents the product of $W_N$ generators
\begin{equation}
W_{-\lambda}^{(j)}=W_{-\lambda_1}^{(j)}\dots W_{-\lambda_k}^{(j)},
\end{equation}
where $k=|\lambda|$, the length of the partition. However, differently from the $N=2$ case, a generic matrix element of descendants operators cannot be written solely in terms of primary matrix elements by using the conformal Ward identities. One class of fields for which this is possible is that of quasi-degenerate fields, for which the conformal weight $\boldsymbol\theta$ is proportional to the weight of the first fundamental representation of $\mathfrak{sl}_N$:
$\boldsymbol\theta=\nu\bs\omega_1$.

The Verma module defined by this highest weight state has $N-2$ null-state decoupling equations, that allow the matrix elements of $V_{\nu\bs\omega_1}$ and its descendants to be expressed in terms of its primary matrix elements
\begin{equation}
\langle\boldsymbol\theta'|V_{\nu\bs\omega_1}|\boldsymbol\theta\rangle\equiv\mathcal{N}(\boldsymbol\theta,\nu\bs\omega_1,\boldsymbol\theta)z^{\Delta_{\boldsymbol\theta'}-\Delta_{\nu\bs\omega_1}-\Delta_{\boldsymbol\theta}},
\end{equation}
where
\begin{equation}
\Delta_{\boldsymbol\theta}=-e_2(\boldsymbol\theta)=\frac{\boldsymbol\theta^2}{2},
\end{equation}
$e_2$ being the second elementary symmetric polynomial in $\theta_1,\dots,\theta_N$.

We will employ also the even more special case of completely degenerate fields, for which $\boldsymbol\theta=\bs{h}_1=\bs\omega_1 $ (first fundamental representation of $\mathfrak{sl}_N$) or $\boldsymbol\theta=-\bs{h}_N=\bs\omega_{N-1}$ (last fundamental representation of $\mathfrak{sl}_N$). In this case there are additional null states that imply further constraints in order for the $\mathcal{N}$'s to be nonvanishing. The fusion of a completely degenerate field with a primary state is
\begin{equation}
V_{\bs{h}_1}|\boldsymbol\theta\rangle = \sum_{s=1}^N\mathcal{N}(\boldsymbol\theta+\bs{h}_s,\bs{h}_1,\boldsymbol\theta)z^{\Delta_{\boldsymbol\theta+\bs{h}_s}-\Delta_{\bs{h}_1}-\Delta_{\boldsymbol\theta}}|\boldsymbol\theta+\bs{h}_s\rangle.
\end{equation}

It turns out to be convenient to restrict the completely degenerate field to a specific fusion channel by using projectors $\mathcal{P}_{\boldsymbol\theta} $:
\begin{align}\label{eq:ReducedDeg}
\phi_{s,\boldsymbol\theta}\equiv\mathcal{P}_{\boldsymbol\theta+\bs{h}_s}V_{\bs{h}_1}\mathcal{P}_{\boldsymbol\theta}, && \bar{\phi}_{s,\boldsymbol\theta}\equiv\mathcal{P}_{\boldsymbol\theta-\bs{h}_s}V_{-\bs{h}_N}\mathcal{P}_{\boldsymbol\theta},
\end{align}
where $s=1,\dots N$. These "reduced" fields have just one fusion channel:
\begin{equation}
\phi_{s,\boldsymbol\theta}|\boldsymbol\theta\rangle=\mathcal{N}(\boldsymbol\theta+\bs{h}_s,\bs{h}_1,\boldsymbol\theta)y^{\Delta_{\boldsymbol\theta+\bs h_s}-\Delta_{\bs{h}_1}-\Delta_{\boldsymbol\theta}}|\boldsymbol\theta+\bs{h}_s\rangle,
\end{equation}
\begin{equation}
\bar{\phi}_{s,\boldsymbol\theta}|\boldsymbol\theta\rangle=\mathcal{N}(\boldsymbol\theta-\bs{h}_s,-\bs{h}_N,\boldsymbol\theta)y^{\Delta_{\boldsymbol\theta-\bs h_s}-\Delta_{\bs{h}_N}-\Delta_{\boldsymbol\theta}}|\boldsymbol\theta-\bs{h}_s\rangle,
\end{equation}
and OPEs
\begin{equation}
\phi_s(z)\bar{\phi}_{s'}(w)\sim\frac{\delta_{s,s'}}{(z-w)^{(N-1)/N}},
\end{equation}
\begin{align}
\phi_s(z)\phi_{s'}(w)\sim 0, && \bar{\phi}_s (z)\bar{\phi}_{s'}(w)\sim 0.
\end{align}
Out of these degenerate fields one can construct N-component free fermions (with very specific $\mathfrak{sl}_N$ charges) like those we have used throughout the body of the paper by the addition of a $U(1)$ boson $\varphi$ satisfying the OPE
\begin{equation}
\varphi(z)\varphi(w)\sim-\frac{1}{2}\log(w-z),
\end{equation}
so that the fields
\begin{align}
\psi_s(z)=e^{i\varphi(z)}\phi_s(z), && \bar{\psi}_s(z)=e^{-i\varphi(z)}\bar{\phi}_s(z)
\end{align}
satisfy the fermion VOA
\begin{equation}
\bar{\psi}_s(z)\psi_{s'}(w)\sim\frac{\delta_{ss'}}{z-w}.
\end{equation}

\end{appendix}
\bibliographystyle{JHEP}
\bibliography{Biblio.bib}

\end{document}